\documentclass[aps,prb,amsmath,amssymb,footinbib,showpacs,twocolumn,superscriptaddress,longbibliography]{revtex4-2}
\usepackage{amsmath}
\usepackage{amssymb}
\usepackage{amsthm}
\usepackage{setspace}
\usepackage{graphicx}
\usepackage{braket}
\usepackage{mathrsfs}
\usepackage{float}
\usepackage[colorlinks = true,linkcolor = red,urlcolor  = blue,citecolor = blue,anchorcolor = blue]{hyperref}
\usepackage[utf8]{inputenc}
\usepackage[english]{babel}
\usepackage{bm}
\usepackage[caption=false]{subfig}
\begin{document}

\title{Topological superfluid phases of attractive Fermi-Hubbard model in narrow-band cold-atom optical lattices}

\author{Tudor D. Stanescu}
\affiliation{Department of Physics and Astronomy, West Virginia University, Morgantown, WV 26506, USA}

\author{Sumanta Tewari}
\affiliation{Department of Physics and Astronomy, Clemson University, Clemson, SC 29634, USA}

\author{V. W. Scarola}
\affiliation{Department of Physics, Virginia Tech, Blacksburg, Virginia 24061, USA}

\begin{abstract}
We investigate the effects of attractive Hubbard interaction on two-component fermionic atoms in narrow two-dimensional (2D) energy bands that exhibit Rashba spin-orbit coupling (SOC) in the presence of an applied Zeeman field. This narrow-band 2D spin-orbit coupled attractive Fermi-Hubbard model can potentially be realized in cold atom systems in optical lattices with artificially engineered Rashba SOC and Zeeman field. Employing a self-consistent mean field approximation for the pairing potential, we uncover a complex phase diagram featuring various topological superfluid (TS) phases, dependent on the chemical potential and the Zeeman field. We focus on the pairing potential and the corresponding quasiparticle gap characterizing the TS phases, which are notably small for a wide-band model with quadratic dispersion near the $\Gamma$-point, as found in earlier work, and we identify the parameter regimes that maximize the gap. We find that, while generally the value of the pairing potential increases with the reduction of the fermionic bandwidth, as expected for narrow- or flat-band systems, the magnitude of the topological gap characterizing the TS phases reaches a maximum of about $10-12.5\%$ of the interaction strength at finite values of the hopping amplitude, Rashba coupling, and Zeeman field.
\end{abstract}

\maketitle

\section{Introduction:} Topologically non-trivial superconductivity \cite{Read2000,Qi2011} or superfluidity has been predicted in two-dimensional (2D) spin-half fermionic systems with Rashba spin-orbit coupling (SOC), Zeeman field, and on-site attractive interaction \cite{Zhang2008, SATO2009a}. The on-site attractive interaction aims to produce s-wave superconductivity or superfluidity in the Rashba system, which becomes topological when the applied Zeeman field exceeds a critical value. Such a spin-orbit coupled Fermi-Hubbard model with an applied Zeeman field can naturally occur in non-centrosymmetric superconductors in a magnetic field \cite{BAUER2012a,Smidman2017} and can potentially be realized in cold atom systems in optical lattices \cite{Zhang2008,SATO2009a,Zhu2011,LIU2014} with artificially engineered SOC \cite{GALITSKI2013a,GOLDMAN2016,ZHANG2018a,VALDES-CURIEL2021} and Zeeman field.  Such a setup would leverage the significant experimental progress in realizing artificial SOC with ultracold fermions \cite{CHEUK2012a,SONG2016,LIVI2016,HUANG2016a,MENG2016a,KOLKOWITZ2017,SONG2019,AEPPLI2022a,LAURIA2022,LIANG2023a}.

A self-consistent mean field calculation for the pairing potential in the spin-orbit coupled Fermi-Hubbard model under a Zeeman field, with fermions following a quadratic dispersion, has shown that although a topological superconducting phase can emerge when the Zeeman field exceeds a critical value, the magnitude of the self-consistent pair potential in the TS phase is extremely small \cite{TEWARI2011}. This makes it challenging to experimentally realize the TS phase in non-centrosymmetric superconductors where the fermion density is low enough to follow a quadratic dispersion. An alternative approach for experimentally achieving a topological superfluid state in spin-orbit coupled systems is to use two-component cold fermionic atoms in optical lattices with artificially engineered SOC and Zeeman field \cite{SATO2009a,Zhu2011,LIU2014,GOLDMAN2016,ZHANG2018a}.  In this setup, an attractive Hubbard-type onsite interaction can be induced by an $s$-wave Feshbach resonance. The fermion density can be sufficiently high so that the (relevant) fermion dispersion relation corresponds to the lattice dispersion, which can be narrowed by adjusting the hopping parameter and the spin-orbit coupling, nearly reaching the flat-band limit \cite{Zhang2013,Lin2014,Chen2016,Hui2017}.

In flat bands, the critical temperature for Cooper pair formation is predicted to be linearly proportional to the attractive interaction between the Cooper pair constituents \cite{Kopnin2011,HEIKKILA2011}. This contrasts sharply with the Bardeen–Cooper–Schrieffer (BCS) theory of superconductivity, valid for quadratic fermion dispersion, where $T_c$
is proportional to the exponential of the inverse of the interaction strength. This indicates that the superconducting critical temperature, and thus the pair potential, is exponentially enhanced in flat-band systems, as compared to dispersive systems within the BCS formalism. The enhancement arises from the high density of states (DOS) and the dominance of interactions over kinetic energy. The band does not need to be perfectly flat to benefit from this; any band where the Hubbard interaction strength is larger than the bandwidth will suffice. In this spirit, we explore in this paper the possibility of realizing topological superfluid states in a narrow-band spin-orbit coupled Fermi-Hubbard system with a Zeeman field, where the relevant system parameters have values smaller than (but comparable to) the interaction strength, so that the pairing potential is enhanced by the narrowness of the band making the realization of robust TS phases more feasible experimentally.

We investigate the effects of attractive Hubbard interactions on two-component fermionic atoms in narrow two-dimensional energy bands with Rashba spin-orbit coupling (SOC) and applied Zeeman field. Using self-consistent mean field theory for evaluating the pairing potential, we uncover a complex phase diagram featuring various topological superfluid phases that depend on the chemical potential and the Zeeman field. We examine whether the self-consistent pairing potential characterizing the TS phases, which is notably small for a simple quadratic dispersion near the $\Gamma$-point, as found in earlier work \cite{TEWARI2011}, increases as the bandwidth narrows in the optical lattice, as predicted for systems near the flat-band limit \cite{Kopnin2011,HEIKKILA2011}. In addition, we determine the (topological) quasiparticle gap that protects different TS phases and identify the optimal system parameters that maximize this gap, ensuring the realization of a robust topological superfluid. We find that, in general, reducing the fermionic bandwidth enhances the pairing potential, consistent with the expected behavior for narrow- or flat-band systems. However, 
the gap characterizing the TS phases does not exceed maximum values evaluated within our mean-field approximation at about $12.5\%$ of the interaction strength. As discussed latter in the work, these maximum gap values are realized within specific parameter regimes characterized by finite values of the hopping amplitude, Rashba coupling, and Zeeman field. A topological gap $10-12\%$ of the attractive interaction strength $V_0$, which can be controlled by a Feshbach resonance, is a significant improvement over the topological gaps potentially achievable in naturally occurring systems, such as noncentrosymmetric superconductors \cite{TEWARI2011}. In recent work \cite{Han2023}, the phase diagram of this model was discussed as a function of Rashba coupling and Zeeman field but only near the half-filled limit. Thus, the TS phase found in this work is the Chern number $C=2$ phase near chemical potential equal to $4t$ (which corresponds to half-filling in our model) in Fig. 2 below. By contrast, in the present work, we discuss a variety of TS phases and discuss the phase diagram as a function of the control parameters chemical potential and Zeeman field. 

The manuscript is organized as follows. In Sec.~\ref{model} we define the attractive Hubbard model and briefly describe our mean-field approach. In Sec.~\ref{results} we present numerical solutions to the mean field equations, discussing the topological phase diagram and the dependence of the pairing potential and quasiparticle gap on relevant parameters and identifying the optimal regimes that maximize the topological gaps characterizing different TS phases. We summarize in Sec.~\ref{conclusion}.

\section{Theoretical model} \label{model}

We consider a two-dimensional (2D) interacting system with Rashba-type spin-orbit coupling (SOC) and perpendicular Zeeman field and we describe it using a tight-binding model defined on a square lattice. In the basis corresponding to the eigenstates $|\phi_{{\bm k}\sigma}\rangle$ of the (non-interacting) system with no SOC and no Zeeman field, the corresponding (second quantized) Hamiltonian has the form $H=H_0 +H_{int}$, with the non-interacting component having the form
\begin{equation}
H_0 = \sum_{{\bm k},\sigma}(\xi_{{\bm k}}+\sigma\Gamma)~\!c_{{\bm k}\sigma}^\dagger c_{{\bm k}\sigma}^{} + \sum_{\bm k}\left(\alpha_{\bm k}~\!c_{{\bm k}\uparrow}^\dagger c_{{\bm k}\downarrow}^{} + h.c.\right),   \label{H0}
\end{equation}
where the operator $c_{{\bm k}\sigma}^\dagger$ creates a particle in the single-particle state $|\phi_{{\bm k}\sigma}\rangle$ with momentum ${\bm k}=(k_x, k_y)$ and spin projection $\sigma =\pm 1\equiv \uparrow\downarrow$ along the (perpendicular) $z$-axis, $\xi_{\bm k} = 2t~\!(2-\cos k_x -\cos k_y) -\mu$ is the energy spectrum relative to the chemical potential $\mu$, assuming nearest-neighbor hopping with amplitude $t$, $\Gamma$ is the Zeeman field, and $\alpha_{\bm k} = \alpha~\! (\sin k_y + i\sin k_x)$ is the SOC contribution. For convenience, we have chosen the lattice constant $a$ as the unit for length, i.e., we have $a=1$. Considering purely local attractive interactions, the second term of the Hamiltonian becomes
\begin{equation}
H_{int} = -V_0\sum_{{\bm k}, {\bm k}^\prime,{\bm q}} c_{{\bm k}+{\bm q}\uparrow}^\dagger c_{{\bm k}\uparrow}^{}~\! c_{{\bm k}^\prime-{\bm q}\downarrow}^\dagger c_{{\bm k}^\prime\downarrow}^{} \label{Hint}
\end{equation}
with $V_0>0$ being the interaction strength. Note that, using a real space basis, $|\phi_{i\sigma}\rangle$, with $i$ labeling lattice sites, the Hamiltonian in Eq. (\ref{Hint}) takes the familiar Hubbard form, $H_{int} = -V_0\sum_i n_{i\uparrow} n_{i\downarrow}$, with $n_{i\sigma} = c_{i\sigma}^\dagger c_{i\sigma}^{}$.  

In this study, we are interested in the narrow band regime characterized by system parameters ($t$, $\alpha$, and $V_0$) and control parameters ($\Gamma$ and $\mu$) having comparable values and, for convenience, we chose the interaction strength, $V_0$ as the energy unit. We focus on investigating the possible emergence of topological superfluid phases and finding the optimal regime(s) associated with these phases. To obtain a clear general understanding of the relevant phases and to efficiently explore the rather large parameter space, we use a mean-field approach. Note, however, that in the narrow band regime correlation effects could be significant, hence testing the validity of our results beyond mean-field remains an important future task. 

Within a mean-field approach, the interaction term can be approximated by a sum of pairing contributions. The corresponding effective Hamiltonian is $H_{eff} = H_0 + H_\Delta$, with the pairing Hamiltonian having the form
\begin{equation}
H_\Delta = \sum_{\bm k} \left(\Delta ~\!c_{{\bm k}\uparrow}^\dagger c_{{-\bm k}\downarrow}^\dagger +  \Delta^* ~\!c_{{-\bm k}\downarrow}^{} c_{{\bm k}\uparrow}^{}\right), \label{HDelta}
\end{equation}
with the pairing potential $\Delta$ being determined self-consistently by solving a mean-field self-consistent gap equation. A derivation of the gap equation can be found, for example, in Ref. \onlinecite{TEWARI2011}. At zero temperature, the pairing $\Delta$ is a solution of the gap equation $\theta(\Delta)=0$, with 
\begin{eqnarray}
\theta(\Delta) &=& -1-\frac{V_0}{4 N_k}\sum_{\bm k}\left[\frac{1}{E_1}+\frac{1}{E_2}\right.
 \label{gapEq} \\
&-&\left.\frac{\Gamma^2}{\sqrt{\xi_{\bm k}^2|\alpha_{\bm k}^{}|^2+\Gamma^2(\xi_{\bm k}^2+|\Delta|^2)}}\left(\frac{1}{E_1}-\frac{1}{E_2}\right)\right], \nonumber
\end{eqnarray}
where the summation is done over the 2D Brillouin zone containing $N_k$ momentum values.  
The energies $E_1\geq 0$ and $E_2>0$, which depend on the system parameters and the pairing $\Delta$, are the (positive) eigenvalues of the effective Hamiltonian, $H_{eff}$. Explicitly, we have
\begin{equation}
E_{1(2)}^2 =  \xi_{\bm k}^2+|\alpha_{\bm k}^{}|^2+\Gamma^2+|\Delta|^2\mp 2\sqrt{\xi_{\bm k}^2|\alpha_{\bm k}^{}|^2\!+\!\Gamma^2(\xi_{\bm k}^2\!+\!|\Delta|^2)}. \label{E12}   
\end{equation}
Note that the solutions of the pairing equation $\theta(\Delta)=0$ corresponding to $d\theta/d\Delta >0$ are unphysical \cite{TEWARI2011}. For a given set of system and control parameters, ($t$, $\alpha$, $V_0$, $\Gamma$, and $\mu$), we determine the pairing $\Delta$ by solving numerically the equation $\theta(\Delta)=0$ and we characterize the low-energy physics of the system (at the mean-field level) by solving the eigenvalue problem associated with the effective Hamiltonian $H_{eff}$. In particular, we calculate the quasiparticle gap $\Delta_{qp}({\bm k}) \leq \Delta$ characterising the system in different regimes. Note that $\Delta_{qp}$ vanishes at a topological quantum phase transition (TQPT), although the pairing $\Delta$ is finite, or in the absence of pairing, $\Delta=0$. 

%%%%%%%%%%%%%%%%%%%%%%%%%%%%%%%%%%%%%%%%%%%%%%%%%%%%%%%%%%%%%%%%
\begin{figure}[t]
\centering
\includegraphics[width=0.45\textwidth]{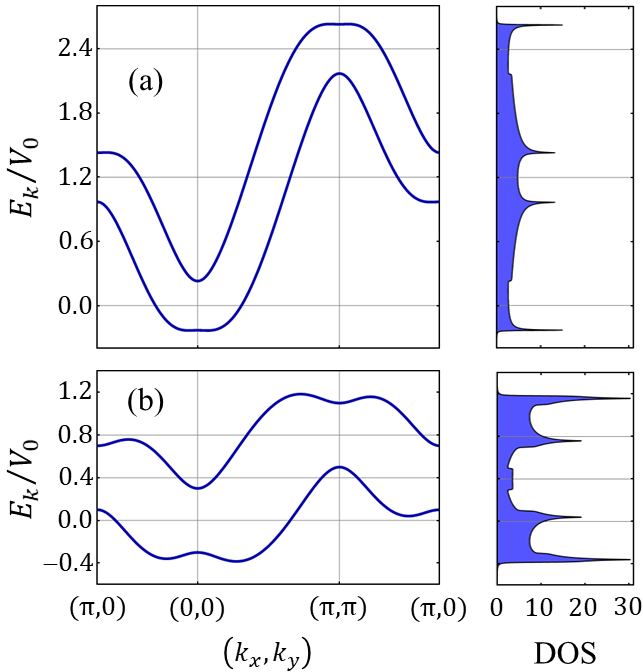}
\caption{Energy spectra (left panels) and the corresponding DOS (right panels) for a non-interacting ($\Delta=0$) narrow-band system with two sets of parameters: (a) $t=0.3V_0$, $\alpha=0.4V_0$, and $\Gamma=0.23V_0$;  (b) $t=0.1V_0$, $\alpha=0.4V_0$, and $\Gamma=0.3V_0$.  Note that these parameters correspond to maximum values of the topological gap for the regimes discussed in Sec.~\ref{results2}. The DOS is given in units of $1/2\pi V_0$, which is the DOS at the bottom of a (spin-degenerate) band for a square lattice model with nearest-neighbor hopping $t=V_0$.}    
\label{Fig0}
\end{figure}
%%%%%%%%%%%%%%%%%%%%%%%%%%%%%%%%%%%%%%%%%%%%%%%%%%%%%%%%%%%%%%%5

The main objective of this study is to determine to what degree narrowing the bandwidth, which enhances the density of states (DOS) and, therefore, is expected to strengthen the pairing potential, results in an enhancement of the topological gap. Two examples of narrow-band non-interacting energy spectra and the corresponding DOS are shown in Fig.~\ref{Fig0}. Note that the density of states is measured in units corresponding to the DOS at the bottom of the band for a square lattice model with nearest-neighbor hopping $t=V_0$, where $V_0$ can be viewed as an unspecified energy scale to be determined (in an interacting system) by the interaction strength. 
In the wide band limit ($t\gg V_0$), e.g., for a wide-band model with quadratic dispersion near the $\Gamma$-point, the corresponding DOS is $1/2\pi t \ll 1/2\pi V_0$.  The DOS shown in Fig.~\ref{Fig0} has values that are orders of magnitude higher that those corresponding to the large bandwidth regime. Below we investigate the impact of this large DOS on the stability of topological superfluid phases that emerge in the presence of onsite attractive interaction.      

\section{Results} \label{results}

In this section we present the results of our numerical analysis, starting with a general discussion of the topological phase diagram for an effective Hamiltonian $H_{eff}$ with constant pairing (Sec. \ref{results1}). We continue in Sec. \ref{results2} with self-consistent calculations of the topological phase diagram corresponding to two different parameter regimes. Finally, in Sec. \ref{results3} we identify the optimal parameter regimes that maximize the topological gap, i.e., the minimum over the Brillouin zone  of $\Delta_{qp}({\bm k})$. 

\subsection{Non-interacting topological phase diagram} \label{results1}

To clearly identify the possible topological phases hosted by the system and to understand the basic structure of the topological phase diagram, we first consider the effective Hamiltonian $H_{eff}=H_0+H_\Delta$, with $H_0$ given by Eq. (\ref{H0}) and a paring term $H_\Delta$ given by Eq. (\ref{HDelta}) having a {constant} (i.e., parameter-independent) pairing potential $\Delta$. Note that $H_{eff}$, which has particle-hole symmetry, but no time-reversal symmetry, belongs to symmetry class D, hence it supports topological phases with a $\mathbb{Z}$ classification in two dimensions \cite{Schnyder2008}. These topological phases are characterized by a Chern number invariant that can be expressed in terms of the Green's function \cite{Volovik1989, Ghosh2010}  $G({\bm k},i\omega) = (i\omega -H_{eff})^{-1}$ as 
\begin{eqnarray}
C=\frac{1}{2}\int\!\!\frac{d^2k}{(2\pi)^2} & &\int \!\!d\omega ~{\rm Tr}\left[G\partial_{k_x}G^{-1}~\! G\partial_{k_y}G^{-1}~\! G\partial_{\omega}G^{-1}      \right. \nonumber \\
\!\!\!\!& &-\left. G\partial_{k_y}G^{-1}~\! G\partial_{k_x}G^{-1}~\! G\partial_{\omega}G^{-1}\right]. \label{Chern}
\end{eqnarray}
We note that the topological phase boundaries separating phases that correspond to different $C$ values are associated with a vanishing energy gap at high-symmetry points in the Brillouin zone, ${\bm K}=(0,0), ~(0,\pi), ~(\pi,0), ~(\pi,\pi)$. For the model considered here, this condition can be expressed analytically using the energy eigenvalues $E_1, E_2$ in Eq.~(5) , with the corresponding phase boundary equations having the form
\begin{eqnarray}
\Gamma^2 = (\mu-\epsilon_i)^2+\Delta^2, \label{GammaB}
\end{eqnarray}
where $\epsilon_1=0$, $\epsilon_2=4t$, and $\epsilon_3=8t$. Note that the topological phase boundaries do not have an explicit dependence on the SOC strength, $\alpha$, but may implicitly depend on this parameter through the pairing potential $\Delta$ (when calculated self-consistently; see Sec. \ref{results2} below).   
%%%%%%%%%%%%%%%%%%%%%%%%%%%%%%%%%%%%%%%%%%%%%%%%%%%%%%%%%%%%%%%%
\begin{figure}[t]
\centering
\includegraphics[width=0.4\textwidth]{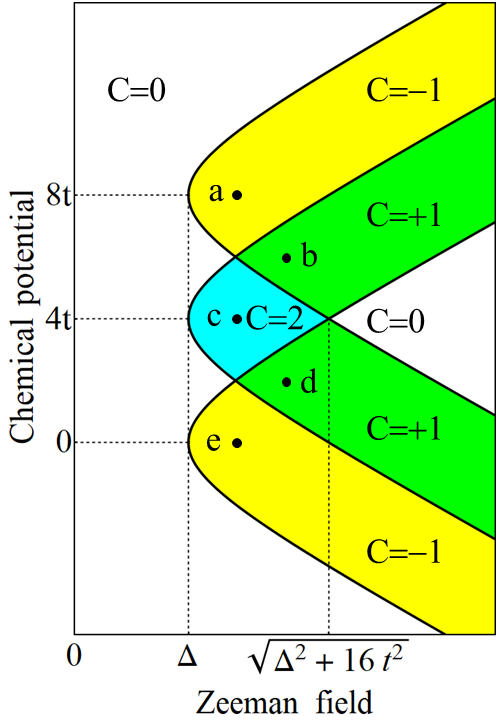}
\caption{Topological phase diagram for a system described by the effective Hamiltonian $H_{eff}=H_)+H_\Delta$ with a constant pairing potential, $\Delta$. The white areas are topologically-trivial, while the colored regions represent three distinct topological superfluid phases. The phase boundaries (black lines) are given by Eq. (\ref{GammaB}), while the values of the Chern topological invariant $C$ are calculated using Eq. (\ref{Chern}). The energy spectra for a finite-width ribbon with control parameter values corresponding to the points $a, b, \dots, e$ are shown in Fig. \ref{Fig2}.}    \label{Fig1}
\end{figure}
%%%%%%%%%%%%%%%%%%%%%%%%%%%%%%%%%%%%%%%%%%%%%%%%%%%%%%%%%%%%%%%5

The topological phase diagram in the Zeeman field--chemical potential plane for a two-dimensional system described by $H_{eff}$ with constant pairing potential $\Delta$ is shown in Fig. \ref{Fig1}. Calculating the topological invariant given by Eq. \ref{Chern} reveals the presence of three topological superfluid phases corresponding to $C=-1$ (yellow shading in Fig. \ref{Fig1}), $C=+1$ (green), and $C=+2$ (cyan). The minimum Zeeman field associated with the emergence of these phases, $\Gamma_{min}=|\Delta|$, is controlled by the pairing potential, while the chemical potential range associated with the topological phases (for a given Zeeman field $\Gamma$) is controlled by the hopping $t$. Note that upon reducing the hopping, $t\rightarrow 0$, the topological regions shrink and eventually collapse.

%%%%%%%%%%%%%%%%%%%%%%%%%%%%%%%%%%%%%%%%%%%%%%%%%%%%%%%%%%%%%%%%
\begin{figure}[t]
\centering
\includegraphics[width=0.44\textwidth]{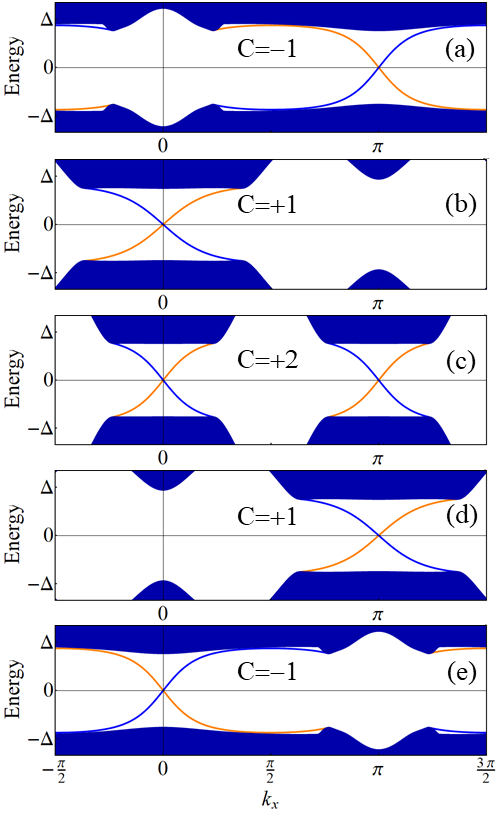}
\caption{Energy spectrum as a function of momentum for $\alpha =4t$ for an infinitely long ribbon with system parameters corresponding to the points marked $a, b, \dots, e$ in Fig. \ref{Fig1}. The dark blue shading indicates bulk states, while the in-gap blue and orange lines correspond to chiral edge modes propagating on the opposite boundaries of the ribbon. Note the opposite chirality of the edge modes supported by the $C=+1$ (green) and $C=-1$ (yellow) phases and the presence of two chiral modes on each edge for $C=+2$ (cyan phase).}    \label{Fig2}
\end{figure}
%%%%%%%%%%%%%%%%%%%%%%%%%%%%%%%%%%%%%%%%%%%%%%%%%%%%%%%%%%%%%%%5

To shed light on the distinction between phases characterized by different (nonzero) values of the Chern invariant, we consider an infinitely-long, finite width ribbon with control parameter values corresponding to the points marked $a, b, \dots, e$ in Fig. \ref{Fig1} and we calculate the energy spectrum as a function of momentum along the ribbon, $k_x$. The results are shown in Fig. \ref{Fig2}. The topological phases are characterized by a finite bulk gap and by the presence of gapless chiral Majorana edge modes. The number and the chirality of these modes are correlated with the value of the Chern invariant characterizing each topological phase, illustrating the bulk-boundary correspondence. In addition, we note the symmetry characterizing the energy spectra in Eq.~(5), which are invariant under the transformation $\mu\rightarrow 8t-\mu$, $k_x\rightarrow \pi-k_x$. For the two-dimensional system, the corresponding symmetry transformation, $\mu\rightarrow 8t-\mu$, ${\bm k}\rightarrow (\pi, \pi)-{\bm k}$, results in $\xi_{\bm k} \rightarrow -\xi_{\bm k}$ and  $\alpha_{\bm k} \rightarrow \alpha_{\bm k}$, so that the energies in Eq. (\ref{E12})  and the function $\theta(\Delta)$ given by Eq. (\ref{gapEq}) remain invariant. Below, we exploit this property and focus on the regime $\mu\leq 4t$.

%%%%%%%%%%%%%%%%%%%%%%%%%%%%%%%%%%%%%%%%%%%%%%%%%%%%%%%%%%%%%%%%
\begin{figure}[t]
\centering
\includegraphics[width=0.44\textwidth]{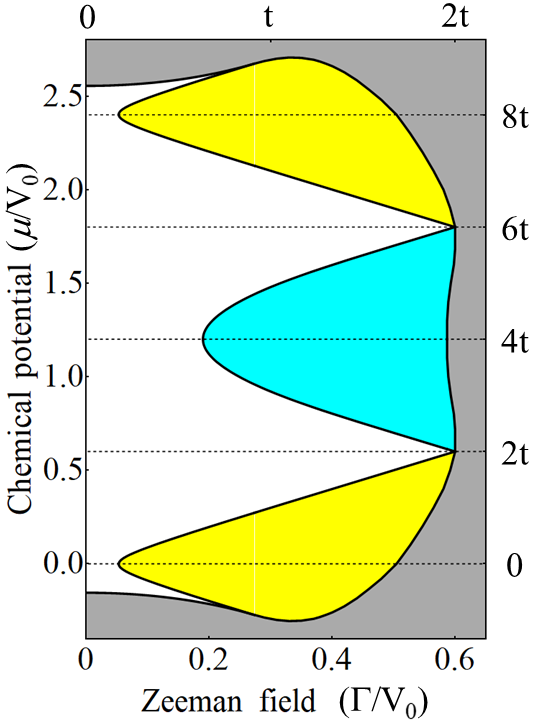}
\caption{Mean-field topological phase diagram for a narrow-band system with $t=0.3V_0$ and $\alpha=0.4V_0$. The light gray area corresponds to an effectively unpaired ($\Delta =0$) phase, while the white, yellow, and cyan regions indicate the same type of gapped phases (trivial or topological) as in Fig. \ref{Fig1}. Note that the green phase ($C=+1$), as seen in Fig.~(2) is not accessible because the hopping parameter $t$ is relatively large.}    \label{Fig3}
\end{figure}
%%%%%%%%%%%%%%%%%%%%%%%%%%%%%%%%%%%%%%%%%%%%%%%%%%%%%%%%%%%%%%%5

\subsection{Mean-field topological phase diagram} \label{results2}

%%%%%%%%%%%%%%%%%%%%%%%%%%%%%%%%%%%%%%%%%%%%%%%%%%%%%%%%%%%%%%%%
\begin{figure}[t]
\centering
\includegraphics[width=0.44\textwidth]{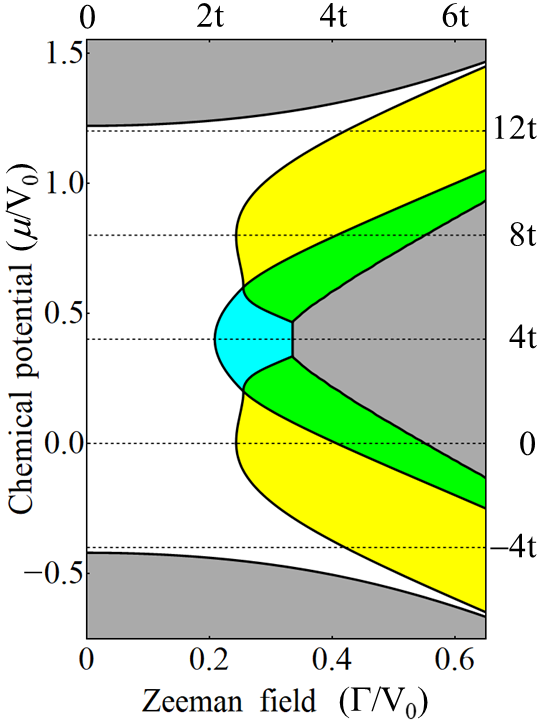}
\caption{Mean-field topological phase diagram for a narrow-band system with $t=0.1V_0$ and $\alpha=0.4V_0$. The color code is the same as in Fig. \ref{Fig1} and Fig. \ref{Fig3}. }    \label{Fig4}
\end{figure}
%%%%%%%%%%%%%%%%%%%%%%%%%%%%%%%%%%%%%%%%%%%%%%%%%%%%%%%%%%%%%%%5

We now consider an interacting system and calculate the topological phase diagram (at the mean-field level) by explicitly determining the pairing potential $\Delta$ as the solution of the equation $\theta(\Delta)=0$ for each set of control parameters, $(\Gamma, \mu)$ for the system parameter values $(t, \alpha) = (0.3, 0.4)V_0$ and   $(t, \alpha) = (0.1, 0.4)V_0$. The corresponding phase diagrams are shown in Fig. \ref{Fig3} and Fig. \ref{Fig4}, respectively. First, we point out that for certain values of the control parameters, the pairing $\Delta$ vanishes. This is in contrast with the wide-band regime with a quadratic dispersion (\cite{TEWARI2011}), where the equation $\theta(\Delta)=0$ has a solution for a nonzero $\Delta$ for all relevant control parameter values and arbitrarily weak interaction strength. This is the result of $E_1$ generating a divergent contribution to the right-hand side of Eq. (\ref{gapEq}) in the limit $|\Delta|\rightarrow 0$ (from $k$ points on the Fermi line \cite{Bardee1957,TEWARI2011}). However, in a narrow-band system the condition $E_1({\bm k})\vert_{\Delta=0}=0$, which implies $|\xi_{\bm k}|=\sqrt{|\alpha_{\bm k}|^2+\Gamma^2}$, cannot be satisfied when the chemical potential is below (or above) the narrow band, or when it lies in a gap between the spin subbands (for large-enough values of $\Gamma$). In these parameter regimes there is no Fermi line and the equation $\theta(\Delta)=0$ does not have a finite solution for arbitrarily weak (attractive) interaction. 
In our numerical calculations, we consider the system as being effectively unpaired (i.e., either a gapless phase or a trivial insulator) if $\Delta < 10^{-4}V_0$. 

%%%%%%%%%%%%%%%%%%%%%%%%%%%%%%%%%%%%%%%%%%%%%%%%%%%%%%%%%%%%%%%%
\begin{figure}[t]
\centering
\includegraphics[width=0.48\textwidth]{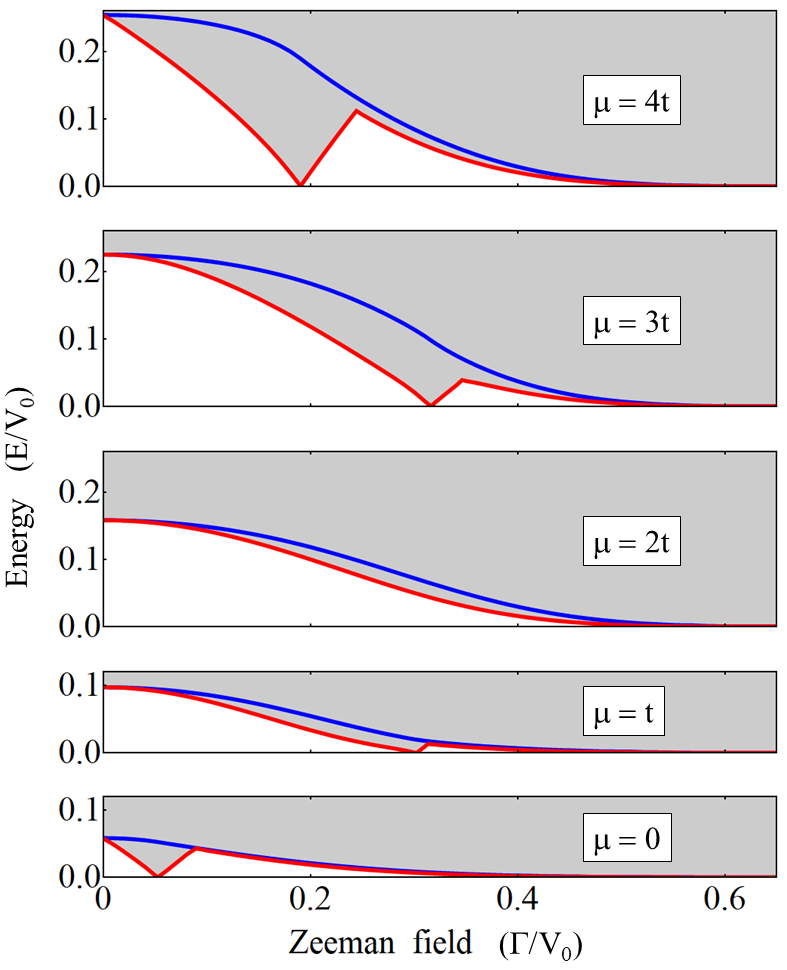}
\caption{Zeeman field dependence of the pairing potential (blue lines) and quasiparticle gap (red lines) for a system with parameters as in Fig. \ref{Fig3} and different chemical potential values. The vanishing of the quasiparticle gap signal a TQPT. The maximum topological gap ($\approx 0.1V_0$) corresponds to half filling ($\mu=4t$, cyan phase with $C=+2$).}    \label{Fig5}
\end{figure}
%%%%%%%%%%%%%%%%%%%%%%%%%%%%%%%%%%%%%%%%%%%%%%%%%%%%%%%%%%%%%%%5

Next, we focus on the topological superfluid phases that emerge at the mean-field level. The phase diagrams in Fig. \ref{Fig3} and Fig. \ref{Fig4} show that these phases can occur within significant control parameter regions, characterized by energy scales of order $V_0$. This is in sharp contrast with a wide-band system with quadratic dispersion \cite{TEWARI2011}, where satisfying the topological condition was found to be quite challenging and the TS phase occurred in a small region of the parameter space. This is one of the central results of the present work: in the narrow-band regime with a lattice dispersion, the topological superfluid phase is much more easily accessible in the space of the control parameters than in the wide-band case with a quadratic dispersion.  However, the extent of a phase in the parameter space, i.e., the area in the phase diagram, provides direct information only about its accessibility for different values of the control parameters, but not about its robustness. The property that is most relevant to the robustness or stability of the topological phase is the size of the topological gap which depends on the pair potential $\Delta$. It has been suggested that the magnitude of $\Delta$ increases with decreasing bandwidth, eventually becoming linearly proportional to the interaction strength $V_0$ in the flat-band limit \cite{Kopnin2011,HEIKKILA2011}. This is in contrast to the BCS theory, which applies to the wide-band regime with a quadratic dispersion. According to the BCS theory the pair potential $\Delta$ depends on the interaction strength $V_0$ as $\Delta \propto e^{-1/g_0V_0}$, where $g_0$ is the density of states at the Fermi energy. Therefore, one can conjecture that going to the narrow- or flat-band regime the magnitude of $\Delta$ should increase exponentially and the topological phase should become more stable. To address this problem, we calculate the quasiparticle gap along five representative $\mu={\rm const.}$ cuts through the phase diagrams. Note that, by symmetry, the energy gaps at chemical potential values $\mu$ and $8t-\mu$ are identical. 

%%%%%%%%%%%%%%%%%%%%%%%%%%%%%%%%%%%%%%%%%%%%%%%%%%%%%%%%%%%%%%%%
\begin{figure}[t]
\centering
\includegraphics[width=0.48\textwidth]{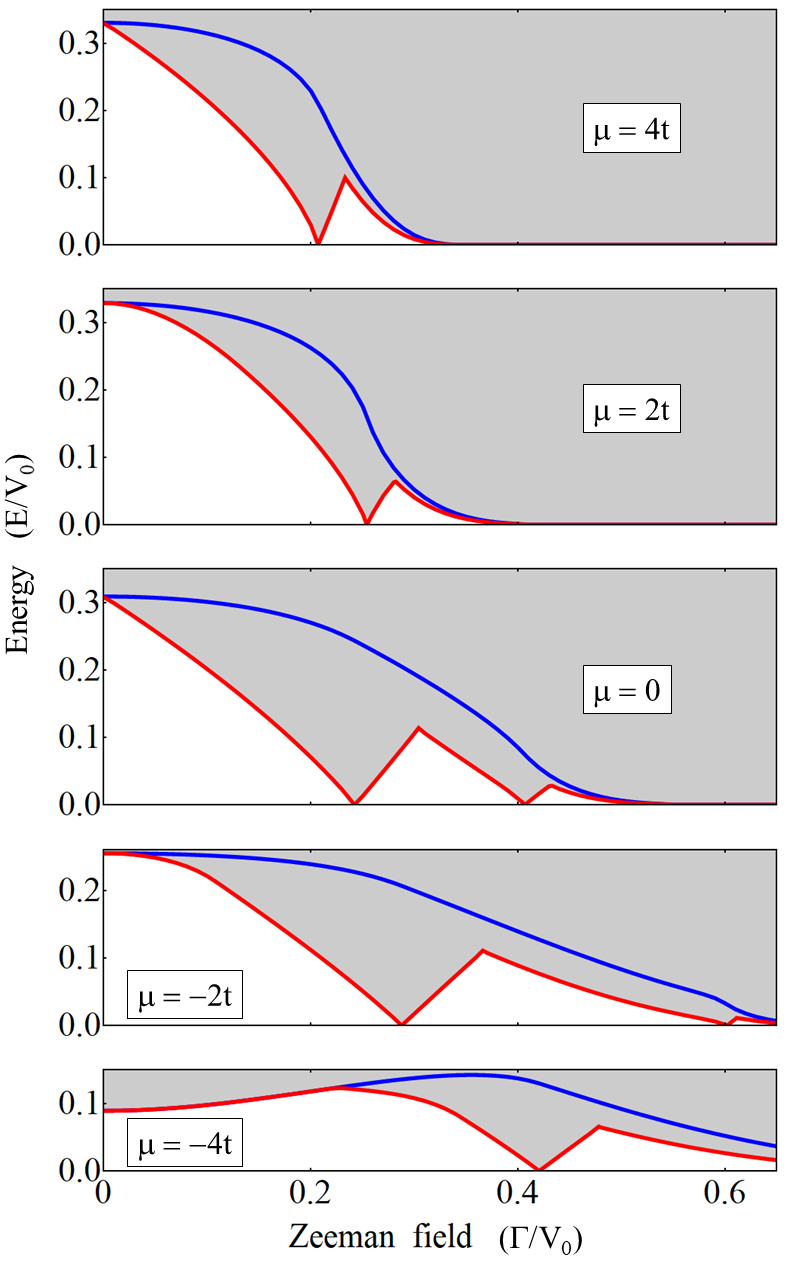}
\caption{Zeeman field dependence of the pairing potential (blue lines) and quasiparticle gap (red lines) for a system with parameters as in Fig. \ref{Fig4} and different chemical potential values. The maximum topological gap values ($\approx 0.1V_0$) correspond to half filling ($\mu=4t$, cyan phase with $C=+2$) and to the yellow ($C=-1$) phase ($\mu =0$ and $\mu=-2t$). Note the non-monotonic dependence of $\Delta$ on the Zeeman field for $\mu=-4t$.}    \label{Fig6}
\end{figure}
%%%%%%%%%%%%%%%%%%%%%%%%%%%%%%%%%%%%%%%%%%%%%%%%%%%%%%%%%%%%%%%5

The results corresponding to the parameter regimes illustrated in Fig. \ref{Fig3} and Fig. \ref{Fig4} are presented in Fig. \ref{Fig5} and Fig. \ref{Fig6}, respectively. The blue lines represent the dependence of the pairing potential $\Delta(\Gamma, \mu)$ on the applied Zeeman field, while the red lines represent the minimum (over the Brillouin zone) of the quasiparticle gap $\Delta_{qp}({\bm k})=E_1({\bm k})$, where $E_1$ is given by Eq. (\ref{E12}) with $\Delta=\Delta(\Gamma, \mu)$. We note that the phase boundary crossings are signaled by the vanishing of the quasiparticle gap. Increasing the Zeeman field typically results in the reduction and eventual collapse of the pairing potential, with the non-monotonic behavior in the lower panel of Fig.\ref{Fig6} ($\mu=-4t$) illustrating the (less likely) scenario associated with small filling values (or, from symmetry, nearly filled systems). The key property that determines the stability of the topological phase is the size of the topological gap, i.e., quasiparticle gap in the topological regime. In the examples shown in Fig. \ref{Fig5} and Fig. \ref{Fig6}, the maximum topological gap values are of the order $0.1V_0$ and are realized in the $C=+2$ (cyan) phase for $\mu=4t$ (i.e., half filling) --- see top panels of Fig. \ref{Fig5} and Fig. \ref{Fig6} --- or in the $C=-1$ (yellow) phase --- panels $\mu=0$ and $\mu=-2t$ in Fig. \ref{Fig6}. By contrast, the values of the topological gap characterizing the $C=+1$ (green) phase from Fig. \ref{Fig4} are significantly smaller, while this phase is completely inaccessible in the regime corresponding to Fig. \ref{Fig3}. Finally, we note that the maximum values of the topological gap are much smaller than the maximum values of the pairing potential, which are realized at $\Gamma=0$, i.e., in the topologically trivial phase. In Fig. \ref{Fig6}, for example, $\Delta(\Gamma=0)$ can exceed $0.3V_0$, which is about three times the maximum value of the topological gap.   

%%%%%%%%%%%%%%%%%%%%%%%%%%%%%%%%%%%%%%%%%%%%%%%%%%%%%%%%%%%%%%%%
\begin{figure}[t]
\centering
\includegraphics[width=0.46\textwidth]{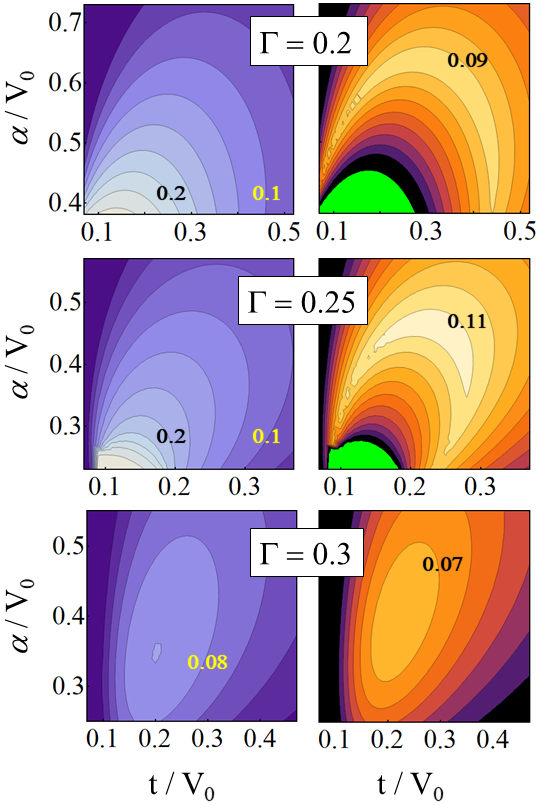}
\caption{Pairing potential (left panels) and topological gap (right panels) as functions of hopping ($t$) and Rashba coupling ($\alpha$) for a half-filled system ($\mu = 4t$, cyan phase). In the right panels the contours are spaced by $0.0125~\!V_0$ (top and middle) and $0.01~\!V_0$ (bottom), while in the right panels the spacing between contours is $0.01~\!V_0$. The green region is topologically trivial. The maximum value of the topological gap is about $0.115~\! V_0$ (for $\Gamma = 0.25V_0$).}    \label{Fig7}
\end{figure}
%%%%%%%%%%%%%%%%%%%%%%%%%%%%%%%%%%%%%%%%%%%%%%%%%%%%%%%%%%%%%%%5

\subsection{Optimization of the system parameters} \label{results3}

Our analysis has shown that a narrow band system described by an attractive Hubbard-type model with spin-orbit coupling and Zeeman field can host topological superfluid phases within significant ranges of control parameters, with maximum values of the topological gap of about $10\%$ of the interaction strength. We can now address the main objective of this work: identifying the optimal parameter regimes that maximize the topological gap and, implicitly, the robustness of the corresponding topological phase. 

%%%%%%%%%%%%%%%%%%%%%%%%%%%%%%%%%%%%%%%%%%%%%%%%%%%%%%%%%%%%%%%%
\begin{figure}[t]
\centering
\includegraphics[width=0.46\textwidth]{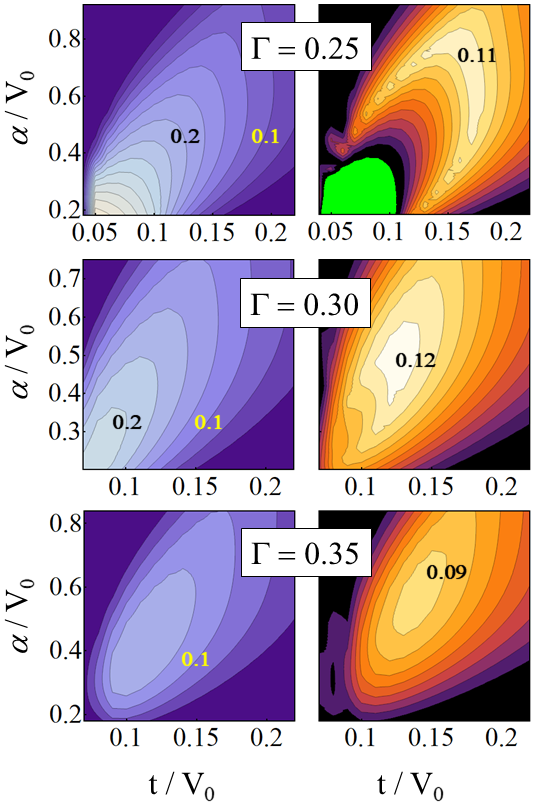}
\caption{Pairing potential (left panels) and topological gap (right panels) as functions of hopping ($t$) and Rashba coupling ($\alpha$) for a system with $\mu = 0$ (yellow phase). The spacing between contours is $0.0125~\!V_0$ (left panels) and $0.01~\! V_0$ (right panels). The maximum topological gap is about $0.125~\!V_0$ (for $\Gamma = 0.3~\! V_0$).}    \label{Fig8}
\end{figure}
%%%%%%%%%%%%%%%%%%%%%%%%%%%%%%%%%%%%%%%%%%%%%%%%%%%%%%%%%%%%%%%5

We emphasize that maximizing the topological gap is not the same as maximizing the pairing potential. As shown by results in Figs. \ref{Fig5} and \ref{Fig6}, the largest values of $\Delta$ are obtained at low Zeeman fields. However, to access a topological phase $\Gamma$ has to exceed the boundary value given by Eq. (\ref{GammaB}),  in particular  one has to satisfy the condition $\Gamma > \Delta$. Hence, the optimal regime cannot be realized at low Zeeman fields. On the other hand, increasing $\Gamma$ eventually results in the collapse of the pairing potential. We conclude that, for a given set of system parameters, the maximum topological gap corresponds to a finite Zeeman field that remains to be determined.

%%%%%%%%%%%%%%%%%%%%%%%%%%%%%%%%%%%%%%%%%%%%%%%%%%%%%%%%%%%%%%%%
\begin{figure}[t]
\centering
\includegraphics[width=0.46\textwidth]{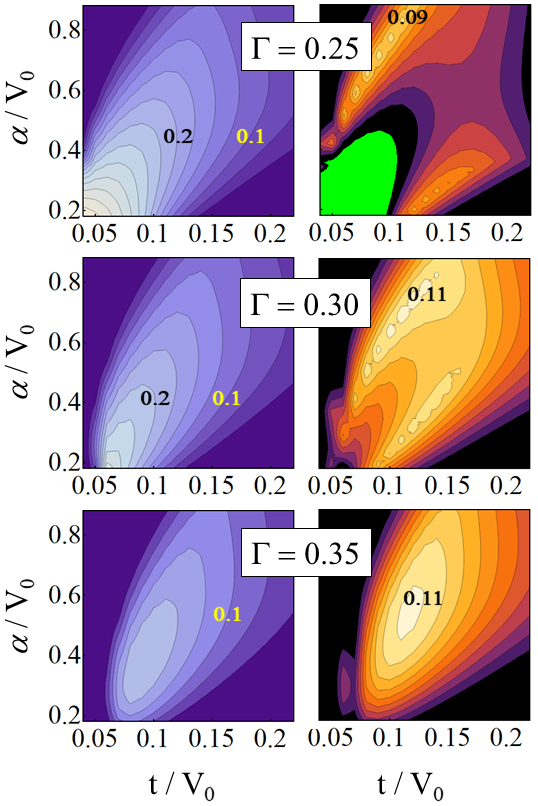}
\caption{Same as in Fig. \ref{Fig8} for a system with chemical potential $\mu=-t$ (yellow phase). Topological gap values above $0.011V_0$ are obtained for Zeeman fields in the range $0.3-0.35~\! V_0$. Note that the pairing potential in the trivial region corresponding to $\Gamma=0.25~\!V_0$ is up to $0.4~\!V_0$ (top left panel).}    \label{Fig9}
\end{figure}
%%%%%%%%%%%%%%%%%%%%%%%%%%%%%%%%%%%%%%%%%%%%%%%%%%%%%%%%%%%%%%%5

Based on our discussion in the previous section, the largest topological gaps are likely to be realized in the $C=+2$ (cyan) or $C=-1$ (yellow) topological phases. For the cyan phase one can easily determine that the gap is maximized at half filling, i.e., for $\mu=4t$. 
For the yellow phase we expect the optimum regime to be away from the phase boundaries, in the range $-3t\lesssim \mu\lesssim t$ or $7t\lesssim \mu\lesssim 11t$. We focus on two specific cases, $\mu=0$ and $\mu=-t$. By symmetry, the same physics will characterize a system with chemical potential values $\mu=8t$ and $\mu=9t$, respectively. Our strategy for identifying the optimal regime is to fix the chemical potential to one of the values mentioned above, consider several Zeeman field strengths, and scan the $t-\alpha$ parameter space. For each set of parameters we first calculate the pairing potential by solving the equation $\theta(\Delta)=0$, then we determine the topological gap by minimizing $\Delta_{qp}({\bm k})$ over the Brillouin zone. 

The results of our numerical analysis are shown in Figs. \ref{Fig7}--\ref{Fig9}. For the half-filled cyan phase (see Fig. \ref{Fig7}), the topological gap is maximized around $\Gamma=0.25~\!V_0$ for hoping parameter values in the range $0.15-0.3~\! V_0$ and Rashba SOC 
strengths of about $0.4-0.45~\!V_0$. For $\Gamma \lesssim 0.25~\!V_0$, the maximum topological gap decreases with decreasing Zeeman field because of the limitations imposed by the topological condition, reaching values of about $\Gamma/2)$, while for $\Gamma \gtrsim 0.25~\!V_0$ the maximum topological gap decreases with increasing Zeeman field, because the pairing potential gets suppressed. This general trend also holds for the yellow topological phase, as illustrated in Fig. \ref{Fig8} and Fig. \ref{Fig9}. In this case, the optimal Zeeman field values are slightly higher (in the range $0.3-0.35~\!V_0$), the optimal hoping is $t\sim 0.12 ~\!V_0$, and the optimal Rashba coupling is in the range $0.5~0.7\!V_0$, while the maximum value of the topological gap is about $V_0/8$. Note that this value is similar to the hopping amplitude.

Our mean-field analysis of the two-dimensional attractive Hubbard-type model with Rashba SOC and Zeeman splitting  in the narrow band limit predicts the emergence of topological superfluid phases with gaps of the order $10\%-12.5\%$ of the interaction strength over a significant range of parameters. The optimal hopping amplitude is $t\approx 0.12-0.25~\!V_0$. Further increasing the hopping, i.e., going toward the wide band regime, reduces the DOS, which results in smaller values of the pairing potential. On the other hand, reducing the hopping results in the collapse of the topological phases, as discussed in Sec. \ref{results1}. The optimal Zeeman field is in the range $0.25-0.35~\!V_0$. Further increasing the field suppresses the pairing, while reducing it imposes stronger constraints on the values of $\Delta$ consistent with the topological condition. The optimal Rashba coupling takes vales in the range $0.4-0.7~\!V_0$. Lowering the SOC strength results in the rapid collapse of the pairing potential at finite Zeeman field, while increasing the SOC strength reduces the maximum pairing (at zero field). Finally, the chemical potential for realizing the optimal $C=+2$ (cyan) phase is $\mu=4t$ (half filling), while the optimal $C=-1$ (yellow) superfluid  phase can be obtained within chemical potential windows of order $0.25~\!V_0$, more specifically $-2t \lesssim \mu\lesssim 0$ and $8t \lesssim \mu\lesssim 10t$.

\section{Conclusion}
\label{conclusion}

We study a model of fermions with attractive Hubbard interaction in a square 2D optical lattice in the presence of SOC and a time-reversal symmetry breaking Zeeman splitting. We are interested in the large Zeeman energy regime, known to drive band flattening \cite{Zhang2013,Lin2014,Chen2016,Hui2017} while simultaneously enabling topological superfluid phases. It has been suggested before that the superconducting/superfluid pairing potential $\Delta$ increases with decreasing bandwidth and in the limit of vanishing bandwidth (flat-band limit) $\Delta$ becomes linearly dependent on the attractive interaction $V_0$. This is in contrast to the case of the wide-band regime with a quadratic dispersion (near the $\Gamma$-point), where the BCS theory applies. According to the BCS theory, the superconducting pair potential $\Delta$ is proportional to the exponential of the inverse of $V_0$, and thus the pair potential should increase exponentially as the bandwidth narrows. This enhancement of the pair potential with narrowing of the bandwidth is ultimately connected to the enhancement of the density of states at the Fermi surface \cite{Kopnin2011, HEIKKILA2011}. In this paper, we explore whether the enhancement of the narrow/flat band density of states could enhance the pairing potential in the TS phase(s) and strengthen the topological superfluid phase by enhancing the corresponding topological gap. 

Using the zeros of the energy eigenvalues (at ${\bm k}=0$), as given by Eq.~(\ref{E12}), and the Chern number in Eq.~(\ref{Chern}), we first show that in the narrow band regime a model with constant pairing potential supports three distinct TS phases that occur in a parameter region significantly expanded as compared to the topological region associated with a wide-band system with quadratic dispersion \cite{TEWARI2011}. 
Since the former applies to optical lattices and the latter applies to naturally-occurring noncentrosymmetric superconductors \cite{TEWARI2011}, we predict that the TS phases should be more easily accessible in optical lattice systems. However, greater accessibility, i.e., an expanded range of control parameters that support TS phases, does not automatically imply more robust topological phases.  In general, the robustness of a TS phase is controlled by the topological (quasiparticle) gap that protects it. Using a self-consistent mean field theory to determine the pairing potential, we calculate the dependence of the topological gap on the relevant system parameters, including the hopping amplitude, Rashba coupling, Zeeman field, and chemical potential. As expected, we find that the pairing potential increases with reducing the bandwidth and, generally, decreases with the applied Zeeman field. However, accessing the topological regime requires a large-enough Zeeman field. In particular,         $\Gamma$ should exceed the value of the (mean-field) pairing potential, $\Delta$. As a result, in the narrow lattice dispersion regime the pairing potential is significantly enhanced as compared to the wide-band, quadratic dispersion regime, but, ultimately, its values in the topological phase are constrained by the Zeeman field $\Gamma$ satisfying the topological condition. In addition, the pairing potential represents an upper bound for the topological gap. To determine the maximum possible values of this gap (within our mean-field approach), we systematically investigate its dependence on the system parameters and identify the optimal regimes. Our best-case scenarios correspond to topological superfluid phases with gaps on the order of $10-12.5\%$ of the attractive interaction strength, $V_0$. This is a significant improvement over the topological gaps achievable in naturally occurring systems, such as noncentrosymmetric superconductors. 

Our work sets the stage to address additional important open questions regarding correlation effects and the thermal stability of topological superfluidity with broken time-reversal symmetry. The effect of correlations on the pairing potential can be investigated using non-perturbative approaches, such as, for example, the dynamical cluster approximation \cite{Doak2023}. On the other hand, the Berenski-Kosterlitz-Thouless transition temperature, $T_{\text{BKT}}$, establishes the temperature below which we can expect to observe topological superfluidity.  $T_{\text{BKT}}$ is proportional to the superfluid stiffness \cite{LIU2014,Xu2015}. In contrast to strictly flat-band systems, the superfluid stiffness is not zero in the present case where the band is narrow but not completely flat. Nonetheless, there should be a contribution to the superfluid stiffness from quantum geometry. Future work will examine the geometric contribution \cite{PEOTTA2015a,Julku2016,LIANG2017,HU2019} to superfluid stiffness in modeling the topological system considered in this paper with broken time-reversal symmetry.

\section{Acknowledgement}
ST and TDS acknowledge support from ONR-N000142312061.  VS acknowledges support from AFOSR (FA2386-21-1-4081, FA9550-23-1-0034, FA9550-19-1-0272). ST and VS acknowledge support from ARO W911NF2210247. 

\bibliography{references}

\begin{thebibliography}{39}%
\makeatletter
\providecommand \@ifxundefined [1]{%
 \@ifx{#1\undefined}
}%
\providecommand \@ifnum [1]{%
 \ifnum #1\expandafter \@firstoftwo
 \else \expandafter \@secondoftwo
 \fi
}%
\providecommand \@ifx [1]{%
 \ifx #1\expandafter \@firstoftwo
 \else \expandafter \@secondoftwo
 \fi
}%
\providecommand \natexlab [1]{#1}%
\providecommand \enquote  [1]{``#1''}%
\providecommand \bibnamefont  [1]{#1}%
\providecommand \bibfnamefont [1]{#1}%
\providecommand \citenamefont [1]{#1}%
\providecommand \href@noop [0]{\@secondoftwo}%
\providecommand \href [0]{\begingroup \@sanitize@url \@href}%
\providecommand \@href[1]{\@@startlink{#1}\@@href}%
\providecommand \@@href[1]{\endgroup#1\@@endlink}%
\providecommand \@sanitize@url [0]{\catcode `\\12\catcode `\$12\catcode `\&12\catcode `\#12\catcode `\^12\catcode `\_12\catcode `\%12\relax}%
\providecommand \@@startlink[1]{}%
\providecommand \@@endlink[0]{}%
\providecommand \url  [0]{\begingroup\@sanitize@url \@url }%
\providecommand \@url [1]{\endgroup\@href {#1}{\urlprefix }}%
\providecommand \urlprefix  [0]{URL }%
\providecommand \Eprint [0]{\href }%
\providecommand \doibase [0]{https://doi.org/}%
\providecommand \selectlanguage [0]{\@gobble}%
\providecommand \bibinfo  [0]{\@secondoftwo}%
\providecommand \bibfield  [0]{\@secondoftwo}%
\providecommand \translation [1]{[#1]}%
\providecommand \BibitemOpen [0]{}%
\providecommand \bibitemStop [0]{}%
\providecommand \bibitemNoStop [0]{.\EOS\space}%
\providecommand \EOS [0]{\spacefactor3000\relax}%
\providecommand \BibitemShut  [1]{\csname bibitem#1\endcsname}%
\let\auto@bib@innerbib\@empty
%</preamble>
\bibitem [{\citenamefont {Read}\ and\ \citenamefont {Green}(2000)}]{Read2000}%
  \BibitemOpen
  \bibfield  {author} {\bibinfo {author} {\bibfnamefont {N.}~\bibnamefont {Read}}\ and\ \bibinfo {author} {\bibfnamefont {D.}~\bibnamefont {Green}},\ }\bibfield  {title} {\bibinfo {title} {Paired states of fermions in two dimensions with breaking of parity and time-reversal symmetries and the fractional quantum hall effect},\ }\href {https://doi.org/10.1103/PhysRevB.61.10267} {\bibfield  {journal} {\bibinfo  {journal} {Phys. Rev. B}\ }\textbf {\bibinfo {volume} {61}},\ \bibinfo {pages} {10267} (\bibinfo {year} {2000})}\BibitemShut {NoStop}%
\bibitem [{\citenamefont {Qi}\ and\ \citenamefont {Zhang}(2011)}]{Qi2011}%
  \BibitemOpen
  \bibfield  {author} {\bibinfo {author} {\bibfnamefont {X.-L.}\ \bibnamefont {Qi}}\ and\ \bibinfo {author} {\bibfnamefont {S.-C.}\ \bibnamefont {Zhang}},\ }\bibfield  {title} {\bibinfo {title} {Topological insulators and superconductors},\ }\href {https://doi.org/10.1103/RevModPhys.83.1057} {\bibfield  {journal} {\bibinfo  {journal} {Rev. Mod. Phys.}\ }\textbf {\bibinfo {volume} {83}},\ \bibinfo {pages} {1057} (\bibinfo {year} {2011})}\BibitemShut {NoStop}%
\bibitem [{\citenamefont {Zhang}\ \emph {et~al.}(2008)\citenamefont {Zhang}, \citenamefont {Tewari}, \citenamefont {Lutchyn},\ and\ \citenamefont {Das~Sarma}}]{Zhang2008}%
  \BibitemOpen
  \bibfield  {author} {\bibinfo {author} {\bibfnamefont {C.}~\bibnamefont {Zhang}}, \bibinfo {author} {\bibfnamefont {S.}~\bibnamefont {Tewari}}, \bibinfo {author} {\bibfnamefont {R.~M.}\ \bibnamefont {Lutchyn}},\ and\ \bibinfo {author} {\bibfnamefont {S.}~\bibnamefont {Das~Sarma}},\ }\bibfield  {title} {\bibinfo {title} {${p}_{x}+i{p}_{y}$ superfluid from $s$-wave interactions of fermionic cold atoms},\ }\href {https://doi.org/10.1103/PhysRevLett.101.160401} {\bibfield  {journal} {\bibinfo  {journal} {Phys. Rev. Lett.}\ }\textbf {\bibinfo {volume} {101}},\ \bibinfo {pages} {160401} (\bibinfo {year} {2008})}\BibitemShut {NoStop}%
\bibitem [{\citenamefont {Sato}\ \emph {et~al.}(2009)\citenamefont {Sato}, \citenamefont {Takahashi},\ and\ \citenamefont {Fujimoto}}]{SATO2009a}%
  \BibitemOpen
  \bibfield  {author} {\bibinfo {author} {\bibfnamefont {M.}~\bibnamefont {Sato}}, \bibinfo {author} {\bibfnamefont {Y.}~\bibnamefont {Takahashi}},\ and\ \bibinfo {author} {\bibfnamefont {S.}~\bibnamefont {Fujimoto}},\ }\bibfield  {title} {\bibinfo {title} {Non-{Abelian} topological order in s-wave superfluids of ultracold fermionic atoms},\ }\href {https://doi.org/10.1103/PhysRevLett.103.020401} {\bibfield  {journal} {\bibinfo  {journal} {Phys. Rev. Lett.}\ }\textbf {\bibinfo {volume} {103}},\ \bibinfo {pages} {020401} (\bibinfo {year} {2009})}\BibitemShut {NoStop}%
\bibitem [{\citenamefont {Bauer}\ and\ \citenamefont {Sigrist}(2012)}]{BAUER2012a}%
  \BibitemOpen
  \bibinfo {editor} {\bibfnamefont {E.}~\bibnamefont {Bauer}}\ and\ \bibinfo {editor} {\bibfnamefont {M.}~\bibnamefont {Sigrist}},\ eds.,\ \href {https://doi.org/10.1007/978-3-642-24624-1} {\emph {\bibinfo {title} {Non-Centrosymmetric Superconductors}}},\ \bibinfo {series} {Lecture {{Notes}} in {{Physics}}}, Vol.\ \bibinfo {volume} {847}\ (\bibinfo  {publisher} {Springer Berlin Heidelberg},\ \bibinfo {address} {Berlin, Heidelberg},\ \bibinfo {year} {2012})\BibitemShut {NoStop}%
\bibitem [{\citenamefont {Smidman}\ \emph {et~al.}(2017)\citenamefont {Smidman}, \citenamefont {Salamon}, \citenamefont {Yuan},\ and\ \citenamefont {Agterberg}}]{Smidman2017}%
  \BibitemOpen
  \bibfield  {author} {\bibinfo {author} {\bibfnamefont {M.}~\bibnamefont {Smidman}}, \bibinfo {author} {\bibfnamefont {M.~B.}\ \bibnamefont {Salamon}}, \bibinfo {author} {\bibfnamefont {H.~Q.}\ \bibnamefont {Yuan}},\ and\ \bibinfo {author} {\bibfnamefont {D.~F.}\ \bibnamefont {Agterberg}},\ }\bibfield  {title} {\bibinfo {title} {Superconductivity and spin–orbit coupling in non-centrosymmetric materials: a review},\ }\href {https://doi.org/10.1088/1361-6633/80/3/036501} {\bibfield  {journal} {\bibinfo  {journal} {Repts. on Prog. in Phys.}\ }\textbf {\bibinfo {volume} {80}},\ \bibinfo {pages} {036501} (\bibinfo {year} {2017})}\BibitemShut {NoStop}%
\bibitem [{\citenamefont {Zhu}\ \emph {et~al.}(2011)\citenamefont {Zhu}, \citenamefont {Shao}, \citenamefont {Wang},\ and\ \citenamefont {Duan}}]{Zhu2011}%
  \BibitemOpen
  \bibfield  {author} {\bibinfo {author} {\bibfnamefont {S.-L.}\ \bibnamefont {Zhu}}, \bibinfo {author} {\bibfnamefont {L.-B.}\ \bibnamefont {Shao}}, \bibinfo {author} {\bibfnamefont {Z.~D.}\ \bibnamefont {Wang}},\ and\ \bibinfo {author} {\bibfnamefont {L.-M.}\ \bibnamefont {Duan}},\ }\bibfield  {title} {\bibinfo {title} {Probing non-abelian statistics of majorana fermions in ultracold atomic superfluid},\ }\href {https://doi.org/10.1103/PhysRevLett.106.100404} {\bibfield  {journal} {\bibinfo  {journal} {Phys. Rev. Lett.}\ }\textbf {\bibinfo {volume} {106}},\ \bibinfo {pages} {100404} (\bibinfo {year} {2011})}\BibitemShut {NoStop}%
\bibitem [{\citenamefont {Liu}\ \emph {et~al.}(2014)\citenamefont {Liu}, \citenamefont {Law},\ and\ \citenamefont {Ng}}]{LIU2014}%
  \BibitemOpen
  \bibfield  {author} {\bibinfo {author} {\bibfnamefont {X.-J.}\ \bibnamefont {Liu}}, \bibinfo {author} {\bibfnamefont {K.~T.}\ \bibnamefont {Law}},\ and\ \bibinfo {author} {\bibfnamefont {T.~K.}\ \bibnamefont {Ng}},\ }\bibfield  {title} {\bibinfo {title} {Realization of {2D} spin-orbit interaction and exotic topological orders in cold atoms},\ }\href {https://doi.org/10.1103/PhysRevLett.112.086401} {\bibfield  {journal} {\bibinfo  {journal} {Phys. Rev. Lett.}\ }\textbf {\bibinfo {volume} {112}},\ \bibinfo {pages} {086401} (\bibinfo {year} {2014})}\BibitemShut {NoStop}%
\bibitem [{\citenamefont {Galitski}\ and\ \citenamefont {Spielman}(2013)}]{GALITSKI2013a}%
  \BibitemOpen
  \bibfield  {author} {\bibinfo {author} {\bibfnamefont {V.}~\bibnamefont {Galitski}}\ and\ \bibinfo {author} {\bibfnamefont {I.~B.}\ \bibnamefont {Spielman}},\ }\bibfield  {title} {\bibinfo {title} {Spin-orbit coupling in quantum gases},\ }\href {https://doi.org/10.1038/nature11841} {\bibfield  {journal} {\bibinfo  {journal} {Nature}\ }\textbf {\bibinfo {volume} {494}},\ \bibinfo {pages} {49} (\bibinfo {year} {2013})}\BibitemShut {NoStop}%
\bibitem [{\citenamefont {Goldman}\ \emph {et~al.}(2016)\citenamefont {Goldman}, \citenamefont {Budich},\ and\ \citenamefont {Zoller}}]{GOLDMAN2016}%
  \BibitemOpen
  \bibfield  {author} {\bibinfo {author} {\bibfnamefont {N.}~\bibnamefont {Goldman}}, \bibinfo {author} {\bibfnamefont {J.~C.}\ \bibnamefont {Budich}},\ and\ \bibinfo {author} {\bibfnamefont {P.}~\bibnamefont {Zoller}},\ }\bibfield  {title} {\bibinfo {title} {Topological quantum matter with ultracold gases in optical lattices},\ }\href {https://doi.org/10.1038/nphys3803} {\bibfield  {journal} {\bibinfo  {journal} {Nat. Phys.}\ }\textbf {\bibinfo {volume} {12}},\ \bibinfo {pages} {639} (\bibinfo {year} {2016})}\BibitemShut {NoStop}%
\bibitem [{\citenamefont {Zhang}\ \emph {et~al.}(2018)\citenamefont {Zhang}, \citenamefont {Zhu}, \citenamefont {Zhao}, \citenamefont {Yan},\ and\ \citenamefont {Zhu}}]{ZHANG2018a}%
  \BibitemOpen
  \bibfield  {author} {\bibinfo {author} {\bibfnamefont {D.-W.}\ \bibnamefont {Zhang}}, \bibinfo {author} {\bibfnamefont {Y.-Q.}\ \bibnamefont {Zhu}}, \bibinfo {author} {\bibfnamefont {Y.~X.}\ \bibnamefont {Zhao}}, \bibinfo {author} {\bibfnamefont {H.}~\bibnamefont {Yan}},\ and\ \bibinfo {author} {\bibfnamefont {S.-L.}\ \bibnamefont {Zhu}},\ }\bibfield  {title} {\bibinfo {title} {Topological quantum matter with cold atoms},\ }\href {https://doi.org/10.1080/00018732.2019.1594094} {\bibfield  {journal} {\bibinfo  {journal} {Adv. in Phys.}\ }\textbf {\bibinfo {volume} {67}},\ \bibinfo {pages} {253} (\bibinfo {year} {2018})}\BibitemShut {NoStop}%
\bibitem [{\citenamefont {{Vald{\'e}s-Curiel}}\ \emph {et~al.}(2021)\citenamefont {{Vald{\'e}s-Curiel}}, \citenamefont {Trypogeorgos}, \citenamefont {Liang}, \citenamefont {Anderson},\ and\ \citenamefont {Spielman}}]{VALDES-CURIEL2021}%
  \BibitemOpen
  \bibfield  {author} {\bibinfo {author} {\bibfnamefont {A.}~\bibnamefont {{Vald{\'e}s-Curiel}}}, \bibinfo {author} {\bibfnamefont {D.}~\bibnamefont {Trypogeorgos}}, \bibinfo {author} {\bibfnamefont {Q.-Y.}\ \bibnamefont {Liang}}, \bibinfo {author} {\bibfnamefont {R.~P.}\ \bibnamefont {Anderson}},\ and\ \bibinfo {author} {\bibfnamefont {I.~B.}\ \bibnamefont {Spielman}},\ }\bibfield  {title} {\bibinfo {title} {Topological features without a lattice in {{Rashba}} spin-orbit coupled atoms},\ }\href {https://doi.org/10.1038/s41467-020-20762-4} {\bibfield  {journal} {\bibinfo  {journal} {Nat. Comm.}\ }\textbf {\bibinfo {volume} {12}},\ \bibinfo {pages} {593} (\bibinfo {year} {2021})}\BibitemShut {NoStop}%
\bibitem [{\citenamefont {Cheuk}\ \emph {et~al.}(2012)\citenamefont {Cheuk}, \citenamefont {Sommer}, \citenamefont {Hadzibabic}, \citenamefont {Yefsah}, \citenamefont {Bakr},\ and\ \citenamefont {Zwierlein}}]{CHEUK2012a}%
  \BibitemOpen
  \bibfield  {author} {\bibinfo {author} {\bibfnamefont {L.~W.}\ \bibnamefont {Cheuk}}, \bibinfo {author} {\bibfnamefont {A.~T.}\ \bibnamefont {Sommer}}, \bibinfo {author} {\bibfnamefont {Z.}~\bibnamefont {Hadzibabic}}, \bibinfo {author} {\bibfnamefont {T.}~\bibnamefont {Yefsah}}, \bibinfo {author} {\bibfnamefont {W.~S.}\ \bibnamefont {Bakr}},\ and\ \bibinfo {author} {\bibfnamefont {M.~W.}\ \bibnamefont {Zwierlein}},\ }\bibfield  {title} {\bibinfo {title} {Spin-injection spectroscopy of a spin-orbit coupled {Fermi} gas},\ }\href {https://doi.org/10.1103/PhysRevLett.109.095302} {\bibfield  {journal} {\bibinfo  {journal} {Phys. Rev. Lett.}\ }\textbf {\bibinfo {volume} {109}},\ \bibinfo {pages} {095302} (\bibinfo {year} {2012})}\BibitemShut {NoStop}%
\bibitem [{\citenamefont {Song}\ \emph {et~al.}(2016)\citenamefont {Song}, \citenamefont {He}, \citenamefont {Zhang}, \citenamefont {Hajiyev}, \citenamefont {Huang}, \citenamefont {Liu},\ and\ \citenamefont {Jo}}]{SONG2016}%
  \BibitemOpen
  \bibfield  {author} {\bibinfo {author} {\bibfnamefont {B.}~\bibnamefont {Song}}, \bibinfo {author} {\bibfnamefont {C.}~\bibnamefont {He}}, \bibinfo {author} {\bibfnamefont {S.}~\bibnamefont {Zhang}}, \bibinfo {author} {\bibfnamefont {E.}~\bibnamefont {Hajiyev}}, \bibinfo {author} {\bibfnamefont {W.}~\bibnamefont {Huang}}, \bibinfo {author} {\bibfnamefont {X.-J.}\ \bibnamefont {Liu}},\ and\ \bibinfo {author} {\bibfnamefont {G.-B.}\ \bibnamefont {Jo}},\ }\bibfield  {title} {\bibinfo {title} {Spin-orbit-coupled two-electron {{Fermi}} gases of ytterbium atoms},\ }\href {https://doi.org/10.1103/PhysRevA.94.061604} {\bibfield  {journal} {\bibinfo  {journal} {Phys. Rev. A}\ }\textbf {\bibinfo {volume} {94}},\ \bibinfo {pages} {061604} (\bibinfo {year} {2016})}\BibitemShut {NoStop}%
\bibitem [{\citenamefont {Livi}\ \emph {et~al.}(2016)\citenamefont {Livi}, \citenamefont {Cappellini}, \citenamefont {Diem}, \citenamefont {Franchi}, \citenamefont {Clivati}, \citenamefont {Frittelli}, \citenamefont {Levi}, \citenamefont {Calonico}, \citenamefont {Catani}, \citenamefont {Inguscio},\ and\ \citenamefont {Fallani}}]{LIVI2016}%
  \BibitemOpen
  \bibfield  {author} {\bibinfo {author} {\bibfnamefont {L.~F.}\ \bibnamefont {Livi}}, \bibinfo {author} {\bibfnamefont {G.}~\bibnamefont {Cappellini}}, \bibinfo {author} {\bibfnamefont {M.}~\bibnamefont {Diem}}, \bibinfo {author} {\bibfnamefont {L.}~\bibnamefont {Franchi}}, \bibinfo {author} {\bibfnamefont {C.}~\bibnamefont {Clivati}}, \bibinfo {author} {\bibfnamefont {M.}~\bibnamefont {Frittelli}}, \bibinfo {author} {\bibfnamefont {F.}~\bibnamefont {Levi}}, \bibinfo {author} {\bibfnamefont {D.}~\bibnamefont {Calonico}}, \bibinfo {author} {\bibfnamefont {J.}~\bibnamefont {Catani}}, \bibinfo {author} {\bibfnamefont {M.}~\bibnamefont {Inguscio}},\ and\ \bibinfo {author} {\bibfnamefont {L.}~\bibnamefont {Fallani}},\ }\bibfield  {title} {\bibinfo {title} {Synthetic dimensions and spin-orbit coupling with an optical clock transition},\ }\href {https://doi.org/10.1103/PhysRevLett.117.220401} {\bibfield  {journal} {\bibinfo  {journal} {Phys. Rev. Lett.}\ }\textbf {\bibinfo {volume} {117}},\ \bibinfo {pages}
  {220401} (\bibinfo {year} {2016})}\BibitemShut {NoStop}%
\bibitem [{\citenamefont {Huang}\ \emph {et~al.}(2016)\citenamefont {Huang}, \citenamefont {Meng}, \citenamefont {Wang}, \citenamefont {Peng}, \citenamefont {Zhang}, \citenamefont {Chen}, \citenamefont {Li}, \citenamefont {Zhou},\ and\ \citenamefont {Zhang}}]{HUANG2016a}%
  \BibitemOpen
  \bibfield  {author} {\bibinfo {author} {\bibfnamefont {L.}~\bibnamefont {Huang}}, \bibinfo {author} {\bibfnamefont {Z.}~\bibnamefont {Meng}}, \bibinfo {author} {\bibfnamefont {P.}~\bibnamefont {Wang}}, \bibinfo {author} {\bibfnamefont {P.}~\bibnamefont {Peng}}, \bibinfo {author} {\bibfnamefont {S.-L.}\ \bibnamefont {Zhang}}, \bibinfo {author} {\bibfnamefont {L.}~\bibnamefont {Chen}}, \bibinfo {author} {\bibfnamefont {D.}~\bibnamefont {Li}}, \bibinfo {author} {\bibfnamefont {Q.}~\bibnamefont {Zhou}},\ and\ \bibinfo {author} {\bibfnamefont {J.}~\bibnamefont {Zhang}},\ }\bibfield  {title} {\bibinfo {title} {Experimental realization of two-dimensional synthetic spin--orbit coupling in ultracold {{Fermi}} gases},\ }\href {https://doi.org/10.1038/nphys3672} {\bibfield  {journal} {\bibinfo  {journal} {Nature Phys}\ }\textbf {\bibinfo {volume} {12}},\ \bibinfo {pages} {540} (\bibinfo {year} {2016})}\BibitemShut {NoStop}%
\bibitem [{\citenamefont {Meng}\ \emph {et~al.}(2016)\citenamefont {Meng}, \citenamefont {Huang}, \citenamefont {Peng}, \citenamefont {Li}, \citenamefont {Chen}, \citenamefont {Xu}, \citenamefont {Zhang}, \citenamefont {Wang},\ and\ \citenamefont {Zhang}}]{MENG2016a}%
  \BibitemOpen
  \bibfield  {author} {\bibinfo {author} {\bibfnamefont {Z.}~\bibnamefont {Meng}}, \bibinfo {author} {\bibfnamefont {L.}~\bibnamefont {Huang}}, \bibinfo {author} {\bibfnamefont {P.}~\bibnamefont {Peng}}, \bibinfo {author} {\bibfnamefont {D.}~\bibnamefont {Li}}, \bibinfo {author} {\bibfnamefont {L.}~\bibnamefont {Chen}}, \bibinfo {author} {\bibfnamefont {Y.}~\bibnamefont {Xu}}, \bibinfo {author} {\bibfnamefont {C.}~\bibnamefont {Zhang}}, \bibinfo {author} {\bibfnamefont {P.}~\bibnamefont {Wang}},\ and\ \bibinfo {author} {\bibfnamefont {J.}~\bibnamefont {Zhang}},\ }\bibfield  {title} {\bibinfo {title} {Experimental observation of a topological band gap opening in ultracold fermi gases with two-dimensional spin-orbit coupling},\ }\href {https://doi.org/10.1103/PhysRevLett.117.235304} {\bibfield  {journal} {\bibinfo  {journal} {Phys. Rev. Lett.}\ }\textbf {\bibinfo {volume} {117}},\ \bibinfo {pages} {235304} (\bibinfo {year} {2016})}\BibitemShut {NoStop}%
\bibitem [{\citenamefont {Kolkowitz}\ \emph {et~al.}(2017)\citenamefont {Kolkowitz}, \citenamefont {Bromley}, \citenamefont {Bothwell}, \citenamefont {Wall}, \citenamefont {Marti}, \citenamefont {Koller}, \citenamefont {Zhang}, \citenamefont {Rey},\ and\ \citenamefont {Ye}}]{KOLKOWITZ2017}%
  \BibitemOpen
  \bibfield  {author} {\bibinfo {author} {\bibfnamefont {S.}~\bibnamefont {Kolkowitz}}, \bibinfo {author} {\bibfnamefont {S.~L.}\ \bibnamefont {Bromley}}, \bibinfo {author} {\bibfnamefont {T.}~\bibnamefont {Bothwell}}, \bibinfo {author} {\bibfnamefont {M.~L.}\ \bibnamefont {Wall}}, \bibinfo {author} {\bibfnamefont {G.~E.}\ \bibnamefont {Marti}}, \bibinfo {author} {\bibfnamefont {A.~P.}\ \bibnamefont {Koller}}, \bibinfo {author} {\bibfnamefont {X.}~\bibnamefont {Zhang}}, \bibinfo {author} {\bibfnamefont {A.~M.}\ \bibnamefont {Rey}},\ and\ \bibinfo {author} {\bibfnamefont {J.}~\bibnamefont {Ye}},\ }\bibfield  {title} {\bibinfo {title} {Spin--orbit-coupled fermions in an optical lattice clock},\ }\href {https://doi.org/10.1038/nature20811} {\bibfield  {journal} {\bibinfo  {journal} {Nature}\ }\textbf {\bibinfo {volume} {542}},\ \bibinfo {pages} {66} (\bibinfo {year} {2017})}\BibitemShut {NoStop}%
\bibitem [{\citenamefont {Song}\ \emph {et~al.}(2019)\citenamefont {Song}, \citenamefont {He}, \citenamefont {Niu}, \citenamefont {Zhang}, \citenamefont {Ren}, \citenamefont {Liu},\ and\ \citenamefont {Jo}}]{SONG2019}%
  \BibitemOpen
  \bibfield  {author} {\bibinfo {author} {\bibfnamefont {B.}~\bibnamefont {Song}}, \bibinfo {author} {\bibfnamefont {C.}~\bibnamefont {He}}, \bibinfo {author} {\bibfnamefont {S.}~\bibnamefont {Niu}}, \bibinfo {author} {\bibfnamefont {L.}~\bibnamefont {Zhang}}, \bibinfo {author} {\bibfnamefont {Z.}~\bibnamefont {Ren}}, \bibinfo {author} {\bibfnamefont {X.-J.}\ \bibnamefont {Liu}},\ and\ \bibinfo {author} {\bibfnamefont {G.-B.}\ \bibnamefont {Jo}},\ }\bibfield  {title} {\bibinfo {title} {Observation of nodal-line semimetal with ultracold fermions in an optical lattice},\ }\href {https://doi.org/10.1038/s41567-019-0564-y} {\bibfield  {journal} {\bibinfo  {journal} {Nat. Phys.}\ }\textbf {\bibinfo {volume} {15}},\ \bibinfo {pages} {911} (\bibinfo {year} {2019})}\BibitemShut {NoStop}%
\bibitem [{\citenamefont {Aeppli}\ \emph {et~al.}(2022)\citenamefont {Aeppli}, \citenamefont {Chu}, \citenamefont {Bothwell}, \citenamefont {Kennedy}, \citenamefont {Kedar}, \citenamefont {He}, \citenamefont {Rey},\ and\ \citenamefont {Ye}}]{AEPPLI2022a}%
  \BibitemOpen
  \bibfield  {author} {\bibinfo {author} {\bibfnamefont {A.}~\bibnamefont {Aeppli}}, \bibinfo {author} {\bibfnamefont {A.}~\bibnamefont {Chu}}, \bibinfo {author} {\bibfnamefont {T.}~\bibnamefont {Bothwell}}, \bibinfo {author} {\bibfnamefont {C.~J.}\ \bibnamefont {Kennedy}}, \bibinfo {author} {\bibfnamefont {D.}~\bibnamefont {Kedar}}, \bibinfo {author} {\bibfnamefont {P.}~\bibnamefont {He}}, \bibinfo {author} {\bibfnamefont {A.~M.}\ \bibnamefont {Rey}},\ and\ \bibinfo {author} {\bibfnamefont {J.}~\bibnamefont {Ye}},\ }\bibfield  {title} {\bibinfo {title} {Hamiltonian engineering of spin-orbit--coupled fermions in a {{Wannier-Stark}} optical lattice clock},\ }\href {https://doi.org/10.1126/sciadv.adc9242} {\bibfield  {journal} {\bibinfo  {journal} {Sci. Adv.}\ }\textbf {\bibinfo {volume} {8}},\ \bibinfo {pages} {eadc9242} (\bibinfo {year} {2022})}\BibitemShut {NoStop}%
\bibitem [{\citenamefont {Lauria}\ \emph {et~al.}(2022)\citenamefont {Lauria}, \citenamefont {Kuo}, \citenamefont {Cooper},\ and\ \citenamefont {Barreiro}}]{LAURIA2022}%
  \BibitemOpen
  \bibfield  {author} {\bibinfo {author} {\bibfnamefont {P.}~\bibnamefont {Lauria}}, \bibinfo {author} {\bibfnamefont {W.-T.}\ \bibnamefont {Kuo}}, \bibinfo {author} {\bibfnamefont {N.~R.}\ \bibnamefont {Cooper}},\ and\ \bibinfo {author} {\bibfnamefont {J.~T.}\ \bibnamefont {Barreiro}},\ }\bibfield  {title} {\bibinfo {title} {Experimental realization of a fermionic spin-momentum lattice},\ }\href {https://doi.org/10.1103/PhysRevLett.128.245301} {\bibfield  {journal} {\bibinfo  {journal} {Phys. Rev. Lett.}\ }\textbf {\bibinfo {volume} {128}},\ \bibinfo {pages} {245301} (\bibinfo {year} {2022})}\BibitemShut {NoStop}%
\bibitem [{\citenamefont {Liang}\ \emph {et~al.}(2023)\citenamefont {Liang}, \citenamefont {Wei}, \citenamefont {Zhang}, \citenamefont {Wang}, \citenamefont {Zhang}, \citenamefont {Wang}, \citenamefont {Qi}, \citenamefont {Liu},\ and\ \citenamefont {Zhang}}]{LIANG2023a}%
  \BibitemOpen
  \bibfield  {author} {\bibinfo {author} {\bibfnamefont {M.-C.}\ \bibnamefont {Liang}}, \bibinfo {author} {\bibfnamefont {Y.-D.}\ \bibnamefont {Wei}}, \bibinfo {author} {\bibfnamefont {L.}~\bibnamefont {Zhang}}, \bibinfo {author} {\bibfnamefont {X.-J.}\ \bibnamefont {Wang}}, \bibinfo {author} {\bibfnamefont {H.}~\bibnamefont {Zhang}}, \bibinfo {author} {\bibfnamefont {W.-W.}\ \bibnamefont {Wang}}, \bibinfo {author} {\bibfnamefont {W.}~\bibnamefont {Qi}}, \bibinfo {author} {\bibfnamefont {X.-J.}\ \bibnamefont {Liu}},\ and\ \bibinfo {author} {\bibfnamefont {X.}~\bibnamefont {Zhang}},\ }\bibfield  {title} {\bibinfo {title} {Realization of {{Qi-Wu-Zhang}} model in spin-orbit-coupled ultracold fermions},\ }\href {https://doi.org/10.1103/PhysRevResearch.5.L012006} {\bibfield  {journal} {\bibinfo  {journal} {Phys. Rev. Research}\ }\textbf {\bibinfo {volume} {5}},\ \bibinfo {pages} {L012006} (\bibinfo {year} {2023})}\BibitemShut {NoStop}%
\bibitem [{\citenamefont {Tewari}\ \emph {et~al.}(2011)\citenamefont {Tewari}, \citenamefont {Stanescu}, \citenamefont {Sau},\ and\ \citenamefont {Das~Sarma}}]{TEWARI2011}%
  \BibitemOpen
  \bibfield  {author} {\bibinfo {author} {\bibfnamefont {S.}~\bibnamefont {Tewari}}, \bibinfo {author} {\bibfnamefont {T.~D.}\ \bibnamefont {Stanescu}}, \bibinfo {author} {\bibfnamefont {J.~D.}\ \bibnamefont {Sau}},\ and\ \bibinfo {author} {\bibfnamefont {S.}~\bibnamefont {Das~Sarma}},\ }\bibfield  {title} {\bibinfo {title} {Topologically non-trivial superconductivity in spin-orbit coupled systems: bulk phases and quantum phase transitions},\ }\href {https://doi.org/10.1088/1367-2630/13/6/065004} {\bibfield  {journal} {\bibinfo  {journal} {New J. Phys.}\ }\textbf {\bibinfo {volume} {13}},\ \bibinfo {pages} {065004} (\bibinfo {year} {2011})}\BibitemShut {NoStop}%
\bibitem [{\citenamefont {Zhang}\ and\ \citenamefont {Zhang}(2013)}]{Zhang2013}%
  \BibitemOpen
  \bibfield  {author} {\bibinfo {author} {\bibfnamefont {Y.}~\bibnamefont {Zhang}}\ and\ \bibinfo {author} {\bibfnamefont {C.}~\bibnamefont {Zhang}},\ }\bibfield  {title} {\bibinfo {title} {Bose-einstein condensates in spin-orbit-coupled optical lattices: Flat bands and superfluidity},\ }\href {https://doi.org/10.1103/PhysRevA.87.023611} {\bibfield  {journal} {\bibinfo  {journal} {Phys. Rev. A}\ }\textbf {\bibinfo {volume} {87}},\ \bibinfo {pages} {023611} (\bibinfo {year} {2013})}\BibitemShut {NoStop}%
\bibitem [{\citenamefont {Lin}\ \emph {et~al.}(2014)\citenamefont {Lin}, \citenamefont {Zhang},\ and\ \citenamefont {Scarola}}]{Lin2014}%
  \BibitemOpen
  \bibfield  {author} {\bibinfo {author} {\bibfnamefont {F.}~\bibnamefont {Lin}}, \bibinfo {author} {\bibfnamefont {C.}~\bibnamefont {Zhang}},\ and\ \bibinfo {author} {\bibfnamefont {V.~W.}\ \bibnamefont {Scarola}},\ }\bibfield  {title} {\bibinfo {title} {Emergent kinetics and fractionalized charge in 1d spin-orbit coupled flatband optical lattices},\ }\href {https://doi.org/10.1103/PhysRevLett.112.110404} {\bibfield  {journal} {\bibinfo  {journal} {Phys. Rev. Lett.}\ }\textbf {\bibinfo {volume} {112}},\ \bibinfo {pages} {110404} (\bibinfo {year} {2014})}\BibitemShut {NoStop}%
\bibitem [{\citenamefont {Chen}\ and\ \citenamefont {Scarola}(2016)}]{Chen2016}%
  \BibitemOpen
  \bibfield  {author} {\bibinfo {author} {\bibfnamefont {M.}~\bibnamefont {Chen}}\ and\ \bibinfo {author} {\bibfnamefont {V.~W.}\ \bibnamefont {Scarola}},\ }\bibfield  {title} {\bibinfo {title} {Stability of emergent kinetics in optical lattices with artificial spin-orbit coupling},\ }\href {https://doi.org/10.1103/PhysRevA.94.043601} {\bibfield  {journal} {\bibinfo  {journal} {Phys. Rev. A}\ }\textbf {\bibinfo {volume} {94}},\ \bibinfo {pages} {043601} (\bibinfo {year} {2016})}\BibitemShut {NoStop}%
\bibitem [{\citenamefont {Hui}\ \emph {et~al.}(2017)\citenamefont {Hui}, \citenamefont {Zhang}, \citenamefont {Zhang},\ and\ \citenamefont {Scarola}}]{Hui2017}%
  \BibitemOpen
  \bibfield  {author} {\bibinfo {author} {\bibfnamefont {H.-Y.}\ \bibnamefont {Hui}}, \bibinfo {author} {\bibfnamefont {Y.}~\bibnamefont {Zhang}}, \bibinfo {author} {\bibfnamefont {C.}~\bibnamefont {Zhang}},\ and\ \bibinfo {author} {\bibfnamefont {V.~W.}\ \bibnamefont {Scarola}},\ }\bibfield  {title} {\bibinfo {title} {Superfluidity in the absence of kinetics in spin-orbit-coupled optical lattices},\ }\href {https://doi.org/10.1103/PhysRevA.95.033603} {\bibfield  {journal} {\bibinfo  {journal} {Phys. Rev. A}\ }\textbf {\bibinfo {volume} {95}},\ \bibinfo {pages} {033603} (\bibinfo {year} {2017})}\BibitemShut {NoStop}%
\bibitem [{\citenamefont {Kopnin}\ \emph {et~al.}(2011)\citenamefont {Kopnin}, \citenamefont {Heikkil\"a},\ and\ \citenamefont {Volovik}}]{Kopnin2011}%
  \BibitemOpen
  \bibfield  {author} {\bibinfo {author} {\bibfnamefont {N.~B.}\ \bibnamefont {Kopnin}}, \bibinfo {author} {\bibfnamefont {T.~T.}\ \bibnamefont {Heikkil\"a}},\ and\ \bibinfo {author} {\bibfnamefont {G.~E.}\ \bibnamefont {Volovik}},\ }\bibfield  {title} {\bibinfo {title} {High-temperature surface superconductivity in topological flat-band systems},\ }\href {https://doi.org/10.1103/PhysRevB.83.220503} {\bibfield  {journal} {\bibinfo  {journal} {Phys. Rev. B}\ }\textbf {\bibinfo {volume} {83}},\ \bibinfo {pages} {220503} (\bibinfo {year} {2011})}\BibitemShut {NoStop}%
\bibitem [{\citenamefont {Heikkil{\"a}}\ \emph {et~al.}(2011)\citenamefont {Heikkil{\"a}}, \citenamefont {Kopnin},\ and\ \citenamefont {Volovik}}]{HEIKKILA2011}%
  \BibitemOpen
  \bibfield  {author} {\bibinfo {author} {\bibfnamefont {T.~T.}\ \bibnamefont {Heikkil{\"a}}}, \bibinfo {author} {\bibfnamefont {N.~B.}\ \bibnamefont {Kopnin}},\ and\ \bibinfo {author} {\bibfnamefont {G.~E.}\ \bibnamefont {Volovik}},\ }\bibfield  {title} {\bibinfo {title} {Flat bands in topological media},\ }\href {https://doi.org/10.1134/S0021364011150045} {\bibfield  {journal} {\bibinfo  {journal} {Jetp Lett.}\ }\textbf {\bibinfo {volume} {94}},\ \bibinfo {pages} {233} (\bibinfo {year} {2011})}\BibitemShut {NoStop}%
\bibitem [{\citenamefont {Han}\ \emph {et~al.}(2023)\citenamefont {Han}, \citenamefont {Yuan},\ and\ \citenamefont {Zhao}}]{Han2023}%
  \BibitemOpen
  \bibfield  {author} {\bibinfo {author} {\bibfnamefont {R.}~\bibnamefont {Han}}, \bibinfo {author} {\bibfnamefont {F.}~\bibnamefont {Yuan}},\ and\ \bibinfo {author} {\bibfnamefont {H.}~\bibnamefont {Zhao}},\ }\bibfield  {title} {\bibinfo {title} {Phase diagram, band structure and density of states in two-dimensional attractive fermi-hubbard model with rashba spin-orbit coupling},\ }\href {https://doi.org/10.1088/1367-2630/acb80d} {\bibfield  {journal} {\bibinfo  {journal} {New Journal of Physics}\ }\textbf {\bibinfo {volume} {25}},\ \bibinfo {pages} {023011} (\bibinfo {year} {2023})}\BibitemShut {NoStop}%
\bibitem [{\citenamefont {Schnyder}\ \emph {et~al.}(2008)\citenamefont {Schnyder}, \citenamefont {Ryu}, \citenamefont {Furusaki},\ and\ \citenamefont {Ludwig}}]{Schnyder2008}%
  \BibitemOpen
  \bibfield  {author} {\bibinfo {author} {\bibfnamefont {A.~P.}\ \bibnamefont {Schnyder}}, \bibinfo {author} {\bibfnamefont {S.}~\bibnamefont {Ryu}}, \bibinfo {author} {\bibfnamefont {A.}~\bibnamefont {Furusaki}},\ and\ \bibinfo {author} {\bibfnamefont {A.~W.~W.}\ \bibnamefont {Ludwig}},\ }\bibfield  {title} {\bibinfo {title} {Classification of topological insulators and superconductors in three spatial dimensions},\ }\href {https://doi.org/10.1103/PhysRevB.78.195125} {\bibfield  {journal} {\bibinfo  {journal} {Phys. Rev. B}\ }\textbf {\bibinfo {volume} {78}},\ \bibinfo {pages} {195125} (\bibinfo {year} {2008})}\BibitemShut {NoStop}%
\bibitem [{\citenamefont {Volovik}\ and\ \citenamefont {Yakovenko}(1989)}]{Volovik1989}%
  \BibitemOpen
  \bibfield  {author} {\bibinfo {author} {\bibfnamefont {G.~E.}\ \bibnamefont {Volovik}}\ and\ \bibinfo {author} {\bibfnamefont {V.~M.}\ \bibnamefont {Yakovenko}},\ }\bibfield  {title} {\bibinfo {title} {Fractional charge, spin and statistics of solitons in superfluid 3he film},\ }\href {https://doi.org/10.1088/0953-8984/1/31/025} {\bibfield  {journal} {\bibinfo  {journal} {Journal of Physics: Condensed Matter}\ }\textbf {\bibinfo {volume} {1}},\ \bibinfo {pages} {5263} (\bibinfo {year} {1989})}\BibitemShut {NoStop}%
\bibitem [{\citenamefont {Ghosh}\ \emph {et~al.}(2010)\citenamefont {Ghosh}, \citenamefont {Sau}, \citenamefont {Tewari},\ and\ \citenamefont {Das~Sarma}}]{Ghosh2010}%
  \BibitemOpen
  \bibfield  {author} {\bibinfo {author} {\bibfnamefont {P.}~\bibnamefont {Ghosh}}, \bibinfo {author} {\bibfnamefont {J.~D.}\ \bibnamefont {Sau}}, \bibinfo {author} {\bibfnamefont {S.}~\bibnamefont {Tewari}},\ and\ \bibinfo {author} {\bibfnamefont {S.}~\bibnamefont {Das~Sarma}},\ }\bibfield  {title} {\bibinfo {title} {Non-abelian topological order in noncentrosymmetric superconductors with broken time-reversal symmetry},\ }\href {https://doi.org/10.1103/PhysRevB.82.184525} {\bibfield  {journal} {\bibinfo  {journal} {Phys. Rev. B}\ }\textbf {\bibinfo {volume} {82}},\ \bibinfo {pages} {184525} (\bibinfo {year} {2010})}\BibitemShut {NoStop}%
\bibitem [{\citenamefont {Doak}\ \emph {et~al.}(2023)\citenamefont {Doak}, \citenamefont {Balduzzi}, \citenamefont {Laurell}, \citenamefont {Dagotto},\ and\ \citenamefont {Maier}}]{Doak2023}%
  \BibitemOpen
  \bibfield  {author} {\bibinfo {author} {\bibfnamefont {P.}~\bibnamefont {Doak}}, \bibinfo {author} {\bibfnamefont {G.}~\bibnamefont {Balduzzi}}, \bibinfo {author} {\bibfnamefont {P.}~\bibnamefont {Laurell}}, \bibinfo {author} {\bibfnamefont {E.}~\bibnamefont {Dagotto}},\ and\ \bibinfo {author} {\bibfnamefont {T.~A.}\ \bibnamefont {Maier}},\ }\bibfield  {title} {\bibinfo {title} {Spin-singlet topological superconductivity in the attractive rashba-hubbard model},\ }\href {https://doi.org/10.1103/PhysRevB.107.224501} {\bibfield  {journal} {\bibinfo  {journal} {Phys. Rev. B}\ }\textbf {\bibinfo {volume} {107}},\ \bibinfo {pages} {224501} (\bibinfo {year} {2023})}\BibitemShut {NoStop}%
\bibitem [{\citenamefont {Xu}\ and\ \citenamefont {Zhang}(2015)}]{Xu2015}%
  \BibitemOpen
  \bibfield  {author} {\bibinfo {author} {\bibfnamefont {Y.}~\bibnamefont {Xu}}\ and\ \bibinfo {author} {\bibfnamefont {C.}~\bibnamefont {Zhang}},\ }\bibfield  {title} {\bibinfo {title} {{Berezinskii-Kosterlitz-Thouless} phase transition in 2d spin-orbit-coupled {Fulde-Ferrell} superfluids},\ }\href {https://doi.org/10.1103/PhysRevLett.114.110401} {\bibfield  {journal} {\bibinfo  {journal} {Phys. Rev. Lett.}\ }\textbf {\bibinfo {volume} {114}},\ \bibinfo {pages} {110401} (\bibinfo {year} {2015})}\BibitemShut {NoStop}%
\bibitem [{\citenamefont {Peotta}\ and\ \citenamefont {T{\"o}rm{\"a}}(2015)}]{PEOTTA2015a}%
  \BibitemOpen
  \bibfield  {author} {\bibinfo {author} {\bibfnamefont {S.}~\bibnamefont {Peotta}}\ and\ \bibinfo {author} {\bibfnamefont {P.}~\bibnamefont {T{\"o}rm{\"a}}},\ }\bibfield  {title} {\bibinfo {title} {Superfluidity in topologically nontrivial flat bands},\ }\href {https://doi.org/10.1038/ncomms9944} {\bibfield  {journal} {\bibinfo  {journal} {Nat. Commun.}\ }\textbf {\bibinfo {volume} {6}},\ \bibinfo {pages} {8944} (\bibinfo {year} {2015})}\BibitemShut {NoStop}%
\bibitem [{\citenamefont {Julku}\ \emph {et~al.}(2016)\citenamefont {Julku}, \citenamefont {Peotta}, \citenamefont {Vanhala}, \citenamefont {Kim},\ and\ \citenamefont {T\"orm\"a}}]{Julku2016}%
  \BibitemOpen
  \bibfield  {author} {\bibinfo {author} {\bibfnamefont {A.}~\bibnamefont {Julku}}, \bibinfo {author} {\bibfnamefont {S.}~\bibnamefont {Peotta}}, \bibinfo {author} {\bibfnamefont {T.~I.}\ \bibnamefont {Vanhala}}, \bibinfo {author} {\bibfnamefont {D.-H.}\ \bibnamefont {Kim}},\ and\ \bibinfo {author} {\bibfnamefont {P.}~\bibnamefont {T\"orm\"a}},\ }\bibfield  {title} {\bibinfo {title} {Geometric origin of superfluidity in the lieb-lattice flat band},\ }\href {https://doi.org/10.1103/PhysRevLett.117.045303} {\bibfield  {journal} {\bibinfo  {journal} {Phys. Rev. Lett.}\ }\textbf {\bibinfo {volume} {117}},\ \bibinfo {pages} {045303} (\bibinfo {year} {2016})}\BibitemShut {NoStop}%
\bibitem [{\citenamefont {Liang}\ \emph {et~al.}(2017)\citenamefont {Liang}, \citenamefont {Vanhala}, \citenamefont {Peotta}, \citenamefont {Siro}, \citenamefont {Harju},\ and\ \citenamefont {T{\"o}rm{\"a}}}]{LIANG2017}%
  \BibitemOpen
  \bibfield  {author} {\bibinfo {author} {\bibfnamefont {L.}~\bibnamefont {Liang}}, \bibinfo {author} {\bibfnamefont {T.~I.}\ \bibnamefont {Vanhala}}, \bibinfo {author} {\bibfnamefont {S.}~\bibnamefont {Peotta}}, \bibinfo {author} {\bibfnamefont {T.}~\bibnamefont {Siro}}, \bibinfo {author} {\bibfnamefont {A.}~\bibnamefont {Harju}},\ and\ \bibinfo {author} {\bibfnamefont {P.}~\bibnamefont {T{\"o}rm{\"a}}},\ }\bibfield  {title} {\bibinfo {title} {Band geometry, {{Berry}} curvature, and superfluid weight},\ }\href {https://doi.org/10.1103/PhysRevB.95.024515} {\bibfield  {journal} {\bibinfo  {journal} {Phys. Rev. B}\ }\textbf {\bibinfo {volume} {95}},\ \bibinfo {pages} {024515} (\bibinfo {year} {2017})}\BibitemShut {NoStop}%
\bibitem [{\citenamefont {Hu}\ \emph {et~al.}(2019)\citenamefont {Hu}, \citenamefont {Hyart}, \citenamefont {Pikulin},\ and\ \citenamefont {Rossi}}]{HU2019}%
  \BibitemOpen
  \bibfield  {author} {\bibinfo {author} {\bibfnamefont {X.}~\bibnamefont {Hu}}, \bibinfo {author} {\bibfnamefont {T.}~\bibnamefont {Hyart}}, \bibinfo {author} {\bibfnamefont {D.~I.}\ \bibnamefont {Pikulin}},\ and\ \bibinfo {author} {\bibfnamefont {E.}~\bibnamefont {Rossi}},\ }\bibfield  {title} {\bibinfo {title} {Geometric and conventional contribution to the superfluid weight in twisted bilayer graphene},\ }\href {https://doi.org/10.1103/PhysRevLett.123.237002} {\bibfield  {journal} {\bibinfo  {journal} {Phys. Rev. Lett.}\ }\textbf {\bibinfo {volume} {123}},\ \bibinfo {pages} {237002} (\bibinfo {year} {2019})}\BibitemShut {NoStop}%
\end{thebibliography}%
%\printbibliography[heading=none]
\end{document}